\documentclass[preprint]{aastex631}

\usepackage{amsmath}

\begin{document}

\title{Primordial non-Gaussianity systematics from redshift mismatch with SPHEREx}

\author[0000-0002-5149-4042]{Chandra Shekhar Saraf}
\affiliation{Korea Astronomy and Space Science Institute, 776 Daedeok-daero, Yuseong-gu, Daejeon 34055, South Korea}

\author[0000-0002-7464-2351]{David Parkinson}
\affiliation{Korea Astronomy and Space Science Institute, 776 Daedeok-daero, Yuseong-gu, Daejeon 34055, South Korea}

\correspondingauthor{Chandra Shekhar Saraf}
\email{cssaraf@kasi.re.kr}



\begin{abstract}

The ability to differentiate between different models of inflation through the imprint of primordial non-Gaussianity (PNG) {requires stringent constraints} on the local PNG parameter $f_{\text{NL}}^{\text{loc}}$. {Upcoming data from the large scale structure survey}s like \textit{Euclid}, Vera C. Rubin Observatory, and the Spectro-Photometer for the History of the Universe, Epoch of Reionization, and Ices Explorer (SPHEREx) {will be instrumental in} advancing our understanding of the inflationary epoch. In this context, we present forecasts on PNG with tomographic {angular power spectra derived from} simulations of SPHEREx. We put forward the effects of redshift bin mismatch of galaxies as a {significant} source of systematic uncertainty in the estimation of both $f_{\text{NL}}^{\text{loc}}$ and galaxy linear halo bias. We simulate $500$ SPHEREx-like galaxy density fields, and divide the galaxies into redshift bins assuming  Gaussian photometric redshift errors. We show that the misclassification of galaxies in redshift bins can result in strong apparent {tensions on $f_{\text{NL}}^{\text{loc}}$ up to $\sim 3-6\sigma$ and up to $\sim 9-12\sigma$ on galaxy bias}. To address this, we propose a scattering matrix formalism that mitigates bin mismatch of galaxies and enables unbiased estimation of cosmological parameters from tomographic angular clustering measurements.

\end{abstract}



\section{Introduction}
Inflation is a widely accepted solution to the horizon and flatness problems, and provides a mechanism for the origin of the density perturbations necessary for the large scale structure (LSS) formation in the Universe \citep{1980PhLB...91...99S,1981MNRAS.195..467S,1981PhRvD..23..347G, 1982PhLB..108..389L, 1983PhLB..129..177L}. The exact dynamics of the inflationary epoch, though, still remains a mystery at large. The current {observations are consistent} with a large number of models for inflation, with different predictions for the inflaton field. Earlier analyses of cosmic microwave background (CMB) and LSS pointed toward a spectrum of nearly Gaussian and scale-invariant primordial fluctuations \citep{2003ApJS..148..119K, 2004PhRvD..69j3501T}. However, many alternative models predict non-Gaussian behaviour of the primordial fluctuations. Primordial non-Gaussianity (PNG) describes deviation from the Gaussian initial density field present after inflation. Most inflationary models suggest that non-Gaussianity depends on the local gravitational potential, and is parameterised by the $f_{\text{NL}}^{\text{loc}}$ parameter:
\begin{equation}
    \Phi(\mathbf{x}) = \phi(\mathbf{x}) + f_{\text{NL}}^{\text{loc}}[\phi^{2}(\mathbf{x})-\langle\phi\rangle^{2}] + \mathcal{O}(\phi^{3}),
\end{equation}
where $\Phi$ is the primordial gravitational potential and $\phi$ is a Gaussian random field \citep{2001PhRvD..63f3002K}.

In general, single field slow roll inflation models predict $|f_{\text{NL}}^{\text{loc}}|\ll 1$ \citep{2003JHEP...05..013M,2004JCAP...10..006C}, whereas multi-field inflationary models can result in stronger non-Gaussianity with $|f_{\text{NL}}^{\text{loc}}|\geq 1$ \citep{2003PhRvD..67b3503L,2004PhRvD..69d3508Z,2004PhR...402..103B}. Thus, to differentiate between single and multi-field inflationary models, a tight constraint on $f_{\text{NL}}^{\text{loc}}$ with at least $\sigma(f_{\text{NL}}^{\text{loc}})\sim 1$ is required.

The current best constraint on $f_{\text{NL}}^{\text{loc}}$ comes from \textit{Planck} bispectrum measurements, $f_{\text{NL}}^{\text{loc}}=-0.9\pm 5.1$ \citep{2020A&A...641A...9P}. Next generation cosmic microwave background (CMB) experiments like CMB-S4 are only expected to improve the \textit{Planck} constraints by a factor of $\sim 2$ \citep{2016arXiv161002743A}. In parallel, tracers of large scale structure (LSS) can be used to constraint PNG in the three-dimensional matter distribution. The current most robust LSS constraints on $f_{\text{NL}}^{\text{loc}}$ come from eBOSS quasars with $\sigma(f_{\text{NL}}^{\text{loc}})=21$ \citep{2022MNRAS.514.3396M}, or from BOSS LRGs with $\sigma(f_{\text{NL}}^{\text{loc}})=28$ \citep{2022PhRvD.106d3506C} and $\sigma(f_{\text{NL}}^{\text{loc}})=31$ \citep{2022arXiv220111518D}. Similar precision on $f_{\text{NL}}^{\text{loc}}$ have also been obtained with photometric galaxy surveys \citep{2014PhRvD..89b3511G,2024MNRAS.532.1902R}. However, systematic errors have been a challenge when estimating $f_{\text{NL}}^{\text{loc}}$ with photometric surveys \citep{2013PASP..125..705P,2014MNRAS.444....2L} or even with spectroscopic surveys \citep{2021MNRAS.506.3439R}.

The constraints on $f_{\text{NL}}^{\text{loc}}$ can be significantly improved by combining information from different LSS tracers \citep{2017MNRAS.466.2780F, 2018MNRAS.479.3490F, 2018PhRvD..97l3540S, 2023EPJC...83..320J, 2023JCAP...08..004S} . One of the avenues to reduce systematic errors is to perform tomographic cross-correlations with surveys like Dark Energy Spectroscopic Instrument (DESI; \citealt{2016arXiv161100036D}), \textit{Euclid} \citep{2024arXiv240513491E}, Vera C. Rubin Observatory Legacy Survey of Space and Time (LSST; \citealt{2009arXiv0912.0201L,2019ApJ...873..111I}), and Spectro-Photometer for the History of the Universe, Epoch of Reionization, and Ices Explorer (SPHEREx; \citealt{2014arXiv1412.4872D}). However, tomographic correlations are inherently plagued by misclassified galaxies in redshift bins due to photometric redshift errors (hereafter, photo-$z$ errors). We identified the impact and mitigation strategy for this redshift bin mismatch of galaxies with simulations of LSST survey in \cite{2024A&A...687A.150S} (hereafter, C24). The main goal of this paper is to forecast the effects of redshift bin mismatch when estimating $f_{\text{NL}}^{\text{loc}}$ from tomographic galaxy angular power spectra with SPHEREx.

SPHEREx\footnote{\url{https://spherex.caltech.edu}} is a NASA medium class space-based observatory that will conduct the first near-infrared spectro-photometric all-sky survey in the wavelength range $0.75< \lambda <5.0\, \mu$m. During its nominal $25$ months mission, SPHEREx will measure the large scale three-dimensional distribution of galaxies. Redshifts for these galaxies will be photometrically determined by fitting templates to spectra, leveraging the nearly-universal $1.6\,\mu$m bump. 
{With the observatory launched in March 2025}, SPHEREx will be the next major experiment to test the theories of inflation. This paper is, then, a timely addition to address the PNG systematics from redshift bin mismatch with SPHEREx.

The paper is arranged as follows: Section \ref{sec:theory} outlines the theoretical background for modelling the angular power spectrum, Section \ref{sec:sims_and_method} describes the simulation setup, methodology for propagation of photo-$z$ errors and estimation of $f_{\text{NL}}^{\text{loc}}$ from power spectra. Our results are presented in Section \ref{sec:results} and Section \ref{sec:conclusions} summarizes our findings and future prospects.

\section{Theory}\label{sec:theory}

Local PNG will increase the amplitude of the halo power spectrum on large scales, resulting in a scale dependent halo bias given by \citep{2008JCAP...08..031S, 2008PhRvD..77l3514D}
\begin{equation}\label{eq:halo_bias_with_fnl}
    b_{h}(k,z) = b_{L}(z) + f_{\text{NL}}^{\text{loc}}[b_{L}(z)-1]\delta_{c}\,\frac{3\Omega_{m}H_{0}^{2}}{k^{2}T(k)D(z)}
\end{equation}
where $b_{L}(z)$ is the linear halo bias and $\delta_{c}$($=1.686$) is the critical over-density for spherical collapse at $z=0$. $\Omega_{m}$ and $H_{0}$ are the matter density parameter and Hubble parameter at $z=0$, $T(k)$ is the linear matter transfer function normalised to $1$ at low $k$, and $D(z)$ is the growth factor. The $f_{\text{NL}}^{\text{loc}}$ parameter can be measured directly from galaxy angular power spectrum or bispectrum when spectroscopic redshifts are available \citep{2013MNRAS.428.1116R, 2020MNRAS.492.1513G, 2021JCAP...05..015M}. However, when spectroscopic data is not available, the galaxy bispectrum cannot be computed reliably. In this case, a tomographic analysis with galaxy auto-power spectra can be used to estimate $f_{\text{NL}}^{\text{loc}}$.

For tomographic analysis, the galaxy angular power spectrum between bins $i$ and $j$ is given by
\begin{equation}\label{eq:galaxy_auto_power_spectrum}
    C_{\ell}^{ij} = \frac{2}{\pi}\int\mathrm{d}k\,k^{2}W^{i}_{\ell}(k)W^{j}_{\ell}(k)P(k)
\end{equation}
where $P(k)$ is the matter power spectrum at redshift $z=0$ computed using \texttt{CAMB} \citep{2000ApJ...538..473L}. $W^{i}_{\ell}(k)$ is the galaxy window function in the $i$th tomographic bin
\begin{equation}
    W^{i}_{\ell}(k) = \int\mathrm{d}z\,\frac{\mathrm{d}N^{i}}{\mathrm{d}z}D(z)b_{h}^{i}(k,z)j_{\ell}(k\chi(z))
\end{equation}
where $\frac{\mathrm{d}N^{i}}{\mathrm{d}z}$ is the redshift distribution of galaxies and $j_{\ell}(k\chi)$ are spherical Bessel functions.
For the range of scales included in the forecast, we used $k_{\text{min}}=0.001\, h/\text{Mpc}$ and $k_{\text{max}}=0.25\, h/\text{Mpc}$. Except $f_{\text{NL}}^{\text{loc}}$, the rest of the cosmological parameters were assumed fixed with values quoted in Table \ref{table:cosmological_parameters}.

\startlongtable
\begin{deluxetable}{ccccc}
    \tablecaption{List of cosmological parameters assumed fixed in our simulations.\label{table:cosmological_parameters}}
    \tablehead{\colhead{$\Omega_{c,0}$} & \colhead{$\Omega_{b,0}$} & \colhead{$H_{0}$} & \colhead{$\sigma_{8}$} & \colhead{$n_{s}$}}
    \startdata 
        $0.265$ & $0.049$ & $67.32$ & $0.811$ & $0.9645$\\
    \enddata
\end{deluxetable}

\section{Simulations and methodology}\label{sec:sims_and_method}

\begin{figure}[ht!]
\gridline{\fig{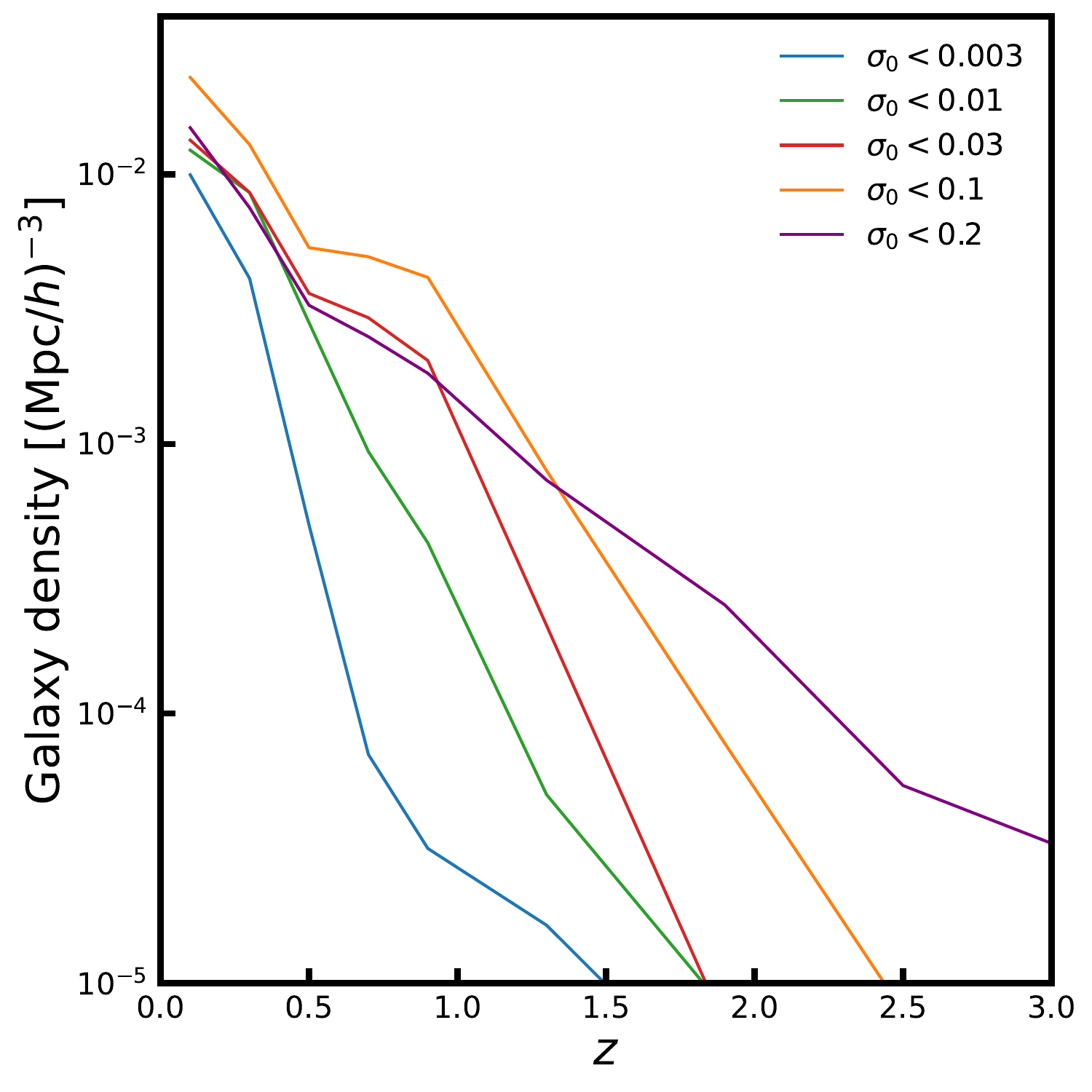}{0.5\textwidth}{(a)}
          \fig{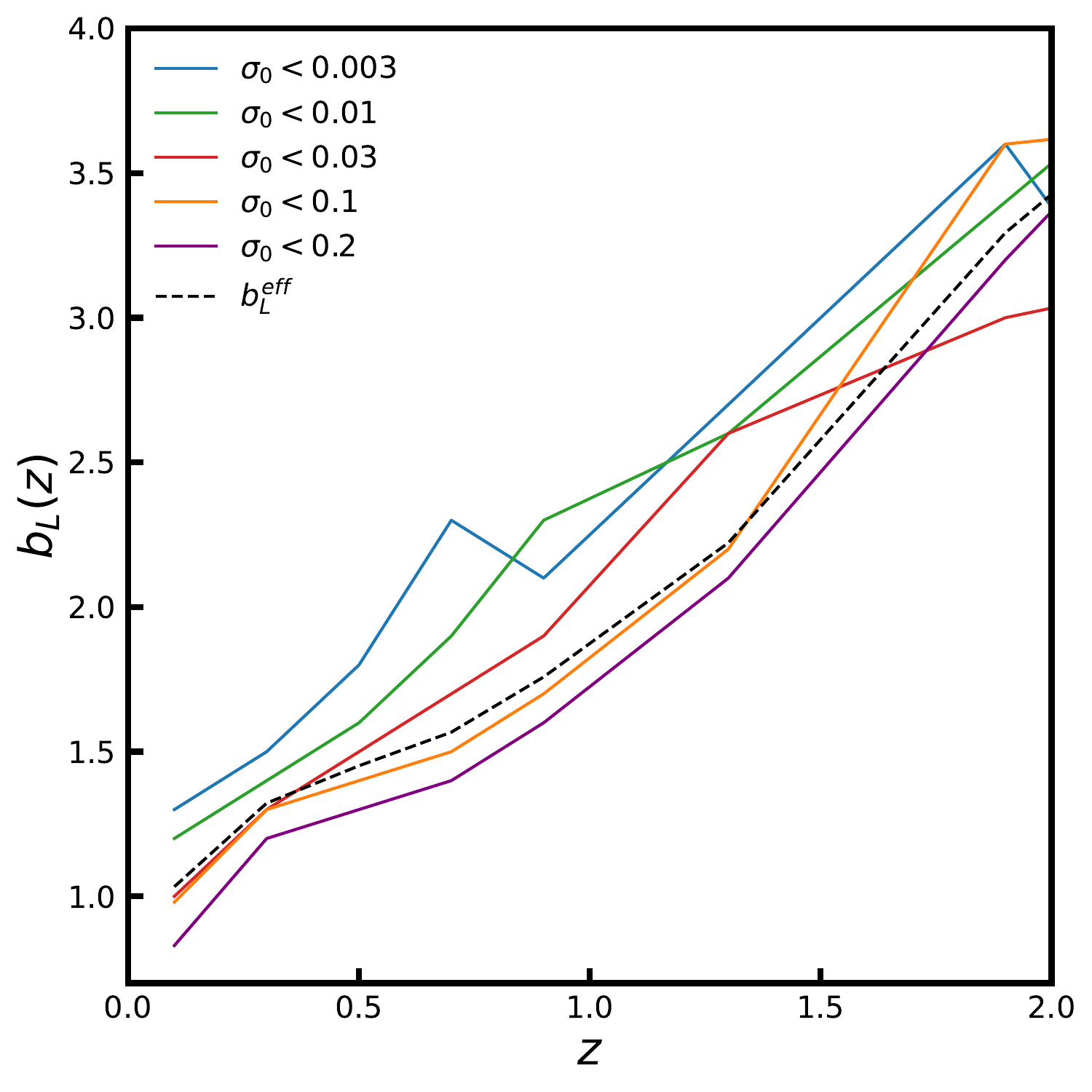}{0.51\textwidth}{(b)}}
\caption{{Fiducial data used in our simulations taken from \cite{2014arXiv1412.4872D}. (a) The comoving number density of galaxies in different redshift accuracy bins. (b) The galaxy linear halo bias in different redshift accuracy bins. The effective galaxy bias is shown with black dashed line.}}
\label{fig:simulation_data_from_olivier_dore_github}
\end{figure}

We used the publicly available code \texttt{GLASS} \citep{2023OJAp....6E..11T} to create $500$ Monte Carlo simulations of lognormal galaxy density field covering $70\%$ of the sky (excluding the galactic plane), and galaxy redshifts $z_{t}$. The photo-$z$s for galaxies $z_{p}$ were generated by drawing a positive random value from Gaussian distribution $\mathcal{N}(z_{t},\sigma_{0}(1+z))$, such that the galaxy number density follows expected SPHEREx redshift accuracy shown in left panel of Figure \ref{fig:simulation_data_from_olivier_dore_github}. The fiducial evolution of linear halo bias used in simulations is shown with black dashed line in the right panel of Figure \ref{fig:simulation_data_from_olivier_dore_github}.

{We created simulations for two configurations of SPHEREx redshift accuracy bins: (i) the three highest-accuracy bins, with $\sigma_0 = 0.003, 0.01, \text{ and } 0.03$, referred to hereafter as Case-I; and (ii) all five redshift accuracy bins, referred to as Case-II. For both configurations, simulations were performed using three different values of the local-type non-Gaussianity parameter: $f_{\text{NL}}^{\text{loc}} = 1$, 10, and 100. In Case-I, each simulated galaxy catalogue was divided into 13 disjoint tomographic bins based on photometric redshifts (photo-$z$), spanning the range $0.0 < z \leq 1.3$ in steps of $\Delta z = 0.1$. In Case-II, we added two additional tomographic bins covering $z = [1.3, 1.5, 2.0]$, resulting in a total of 15 disjoint redshift bins.}

We built galaxy over-density maps from photometric number count maps at HEALPix \citep{2005ApJ...622..759G} resolution $N_{\text{side}}=256$ using
\begin{equation}
    g(\hat{{\textbf{n}}}) = \frac{n(\hat{{\textbf{n}}})-\overline{n}}{\overline{n}},
    \label{eq:gal_overdensity}
\end{equation}
where $n(\hat{{\textbf{n}}})$ is the number of galaxies at angular position $\hat{{\textbf{n}}}$, and $\overline{n}$ is the mean number of galaxies per pixel.

\subsection{Estimating the power spectra}
We used the \texttt{MASTER} algorithm \citep{2002ApJ...567....2H} implemented in \texttt{NaMASTER} \citep{2019MNRAS.484.4127A} to compute the full sky power spectra in every tomographic bin. We binned the power spectra in bins of $\Delta\ell=10$ from $\ell=2$ to $\ell=80$. We computed the sample covariance matrix in each tomographic bin from $500$ simulations using
\begin{equation}\label{eq:sample_covariance}
    \mathbf{K}_{\ell\ell'}^{gg,gg} = \frac{1}{\text{N}_{\text{s}}-1}\sum_{i=1}^{\text{N}_{\text{s}}}\left(\tilde{C}^{gg,i}_\ell-\left\langle \tilde{C}_{\ell}^{gg} \right \rangle\right)\left(\tilde{C}^{gg,i}_\ell-\left\langle  \tilde{C}_{\ell}^{gg}\right\rangle\right),
\end{equation}
where $\text{N}_{\text{s}}$ is the total number of simulations, $\tilde{C}^{gg,i}_\ell$ is the power spectrum estimated from the $i\text{th}$ simulation and
\begin{equation}\label{eq:mock_avg}
    \left\langle  \tilde{C}_{\ell}^{gg}\right\rangle = \frac{1}{\text{N}_{\text{s}}}\sum_{i=1}^{\text{N}_{\text{s}}}\tilde{C}^{gg,i}_\ell
\end{equation}

\subsection{Propagation of photo-z errors}\label{sec:propagate_photoz_errors}
\begin{figure}[ht!]
\gridline{\fig{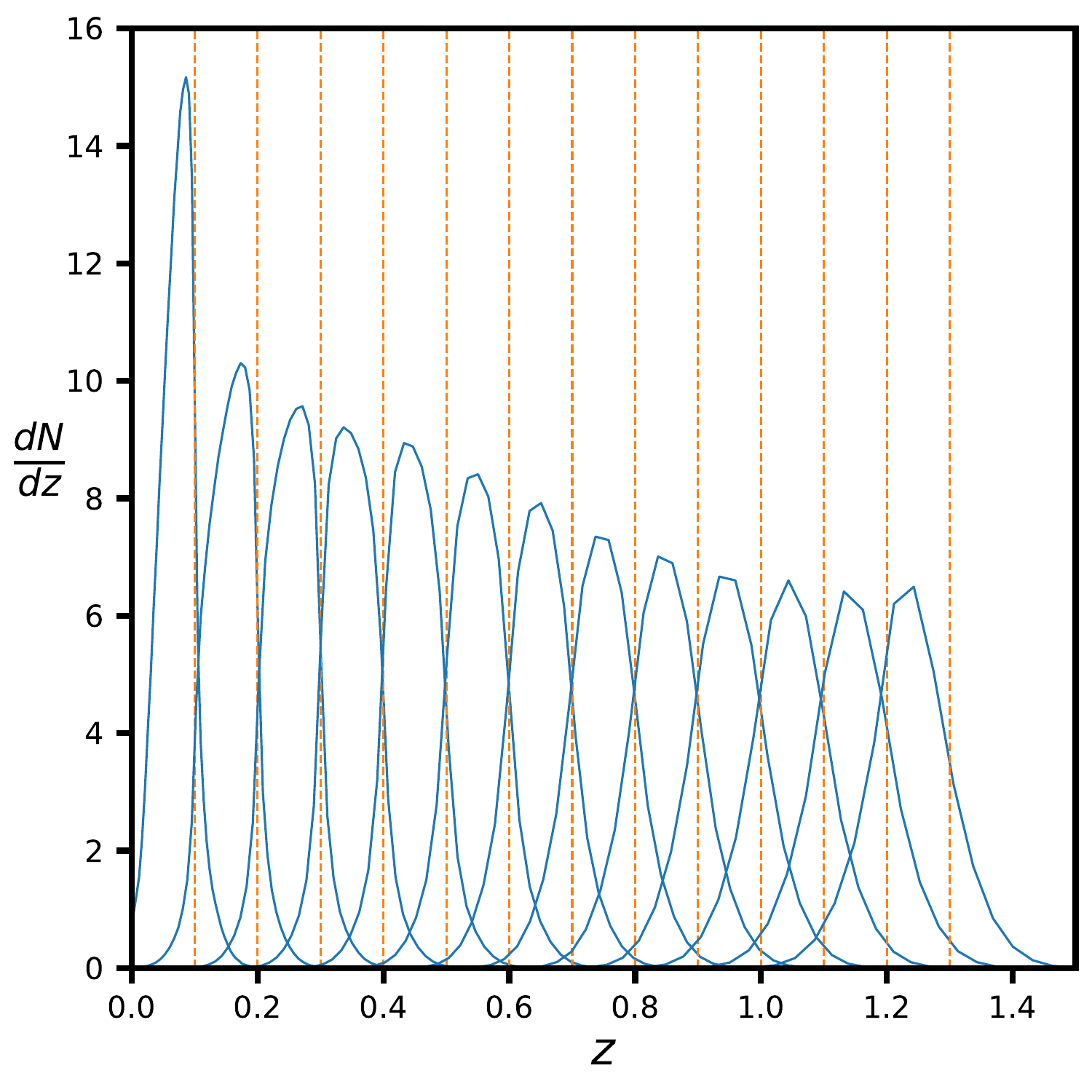}{0.5\textwidth}{(a) Case-I}
          \fig{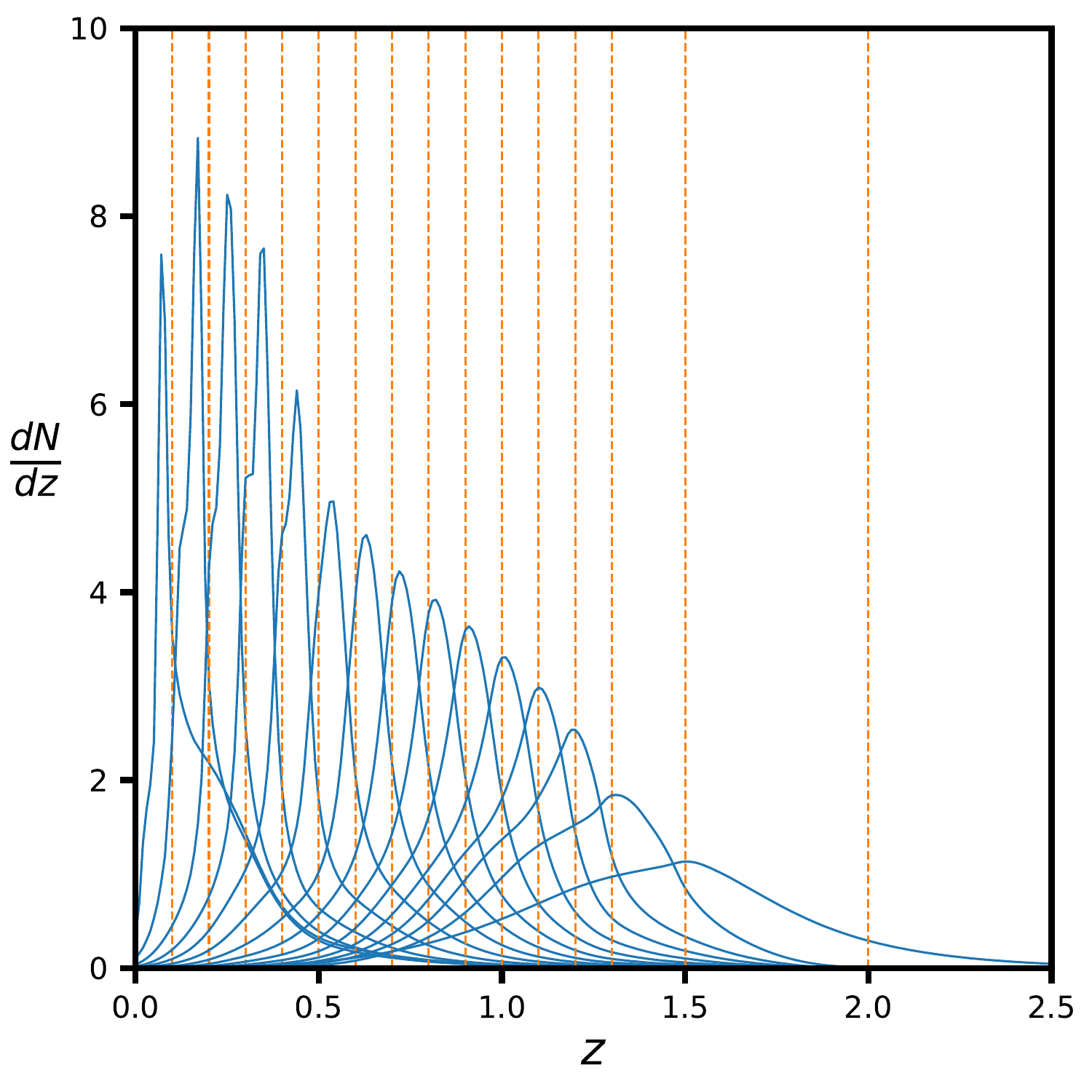}{0.5\textwidth}{(b) Case-II}}
\caption{Effect of photo-z scatter on tomographic redshift distributions for (a) Case-I and (b) Case-II. The dashed orange lines mark the boundaries of redshift bins. The blue solid curves are the redshift distributions obtained after convolution.}
\label{fig:tomographic_galaxy_redshift_distribution}
\end{figure}

We accounted for photo-z uncertainties in our analysis by convoluting the photometric redshift distribution with the conditional probability $p(z_{t}-z_{p}|z_{p})$, which we call the photo-z error distribution. An estimate of the true redshift distribution for $i$th tomographic bin can then be given as
\begin{equation}
    \frac{\mathrm{d}N^{i}(z)}{\mathrm{d}z} = \int\mathrm{d}z_{p}\frac{\mathrm{d}N(z_{p})}{\mathrm{d}z_{p}}\Theta^{i}(z_{p})p^{i}(z-z_{p}|z_{p}),
    \label{eq:true_dist_conv}
\end{equation}
where $\frac{\mathrm{d}N(z_{p})}{\mathrm{d}z_{p}}$ is the observed photometric redshift distribution of galaxies, and $\Theta^{i}(z_{p})$ is a step function defining the $i$th redshift bin,
\begin{equation}
    \Theta^{i}(z) = \begin{cases}
        1,& \text{if}\quad z^{i}_{\text{min}}\leq z<z^{i+1}_{\text{min}}\\
        0,& \text{otherwise}.
    \end{cases}
    \label{eq:window_function}
\end{equation}
{The true galaxy redshift distributions, obtained after convolving with photometric redshift errors, are shown in Figure \ref{fig:tomographic_galaxy_redshift_distribution}. The blue curves illustrate how the initially disjoint redshift bins are broadened due to photo-$z$ uncertainties. The dashed vertical orange lines indicate the boundaries of the redshift bins. A marked increase in the overlap between tomographic redshift distributions is evident when comparing Case-I (Figure \ref{fig:tomographic_galaxy_redshift_distribution}a) to Case-II (Figure \ref{fig:tomographic_galaxy_redshift_distribution}b).}

\subsection{Parameter estimation}
To forecast constraints on $f_{\text{NL}}^{\text{loc}}$ parameter we used the maximum likelihood estimation method. The log-likelihood function takes the form
\begin{equation}
\begin{split}
	\log \mathcal{L} &= -\frac{1}{2}[\mathbf{d}_{\ell}-\mathbf{t}_{\ell}(\theta)]^\top(\mathbf{K}_{\ell\ell'})^{-1}[\mathbf{d}_{\ell}-\mathbf{t}_{\ell}(\theta)],
\end{split}
\label{eq:joint_likeli}
\end{equation}
where $\mathbf{d}_{\ell} = \{\left\langle  \tilde{C}_{\ell}^{gg}\right\rangle\}$ is the joint data vector from $13$ redshift bins, $\mathbf{t}_{\ell}(\theta)$ is the theory vector, $\theta$ represents the free parameters set ($13$ galaxy bias parameters + $f_{\text{NL}}^{\text{loc}}$), and $\mathbf{K}_{\ell\ell'}$ is the joint covariance matrix given by
\begin{equation}
    \mathbf{K}_{\ell\ell'} = 
    \begin{bmatrix} 
        \mathbf{K}_{\ell\ell'}^{g_{1}g_{1},g_{1}g_{1}} & \mathbf{K}_{\ell\ell'}^{g_{1}g_{1},g_{2}g_{2}} & \dots & \mathbf{K}_{\ell\ell'}^{g_{1}g_{1},g_{13}g_{13}}\\
        \mathbf{K}_{\ell\ell'}^{g_{2}g_{2},g_{1}g_{1}} & \mathbf{K}_{\ell\ell'}^{g_{2}g_{2},g_{2}g_{2}} & \dots & \mathbf{K}_{\ell\ell'}^{g_{2}g_{2},g_{13}g_{13}}\\
        \vdots & \vdots & \ddots & \vdots \\
        \mathbf{K}_{\ell\ell'}^{g_{13}g_{13},g_{1}g_{1}} & \mathbf{K}_{\ell\ell'}^{g_{13}g_{13},g_{2}g_{2}} & \dots & \mathbf{K}_{\ell\ell'}^{g_{13}g_{13},g_{13}g_{13}}.
    \end{bmatrix}
\end{equation}
We used flat priors for parameters in the range ${b_{L} \in [0,20]}$ and ${f_{\text{NL}}^{\text{loc}} \in [-100,200]}$. We used the \texttt{EMCEE} package \citep{2013PASP..125..306F} to effectively sample the parameter space.

\section{Results}\label{sec:results}

\subsection{Pipeline validation without redshift errors}
Before discussing the effects of photo-$z$ errors on parameters, it is worth to validate our analysis pipeline. We did this by estimating $f_{\text{NL}}^{\text{loc}}$ and galaxy bias from simulations before adding photo-$z$ errors. {The recovered values of $f_{\text{NL}}^{\text{loc}}$ parameter for Case-I are shown in Figure \ref{fig:comp_fnl_wout_photoz_errros}. We find similar results for Case-II simulations.} We note that there are no intrinsic systematics in our analysis pipeline. In the following sections, we present the effects of photo-$z$ errors on $f_{\text{NL}}^{\text{loc}}$ and galaxy bias.

\subsection{Constraints on \texorpdfstring{$f_{\text{NL}}^{\text{loc}}$}{Lg} and galaxy bias}

\begin{figure}[ht!]
\gridline{\fig{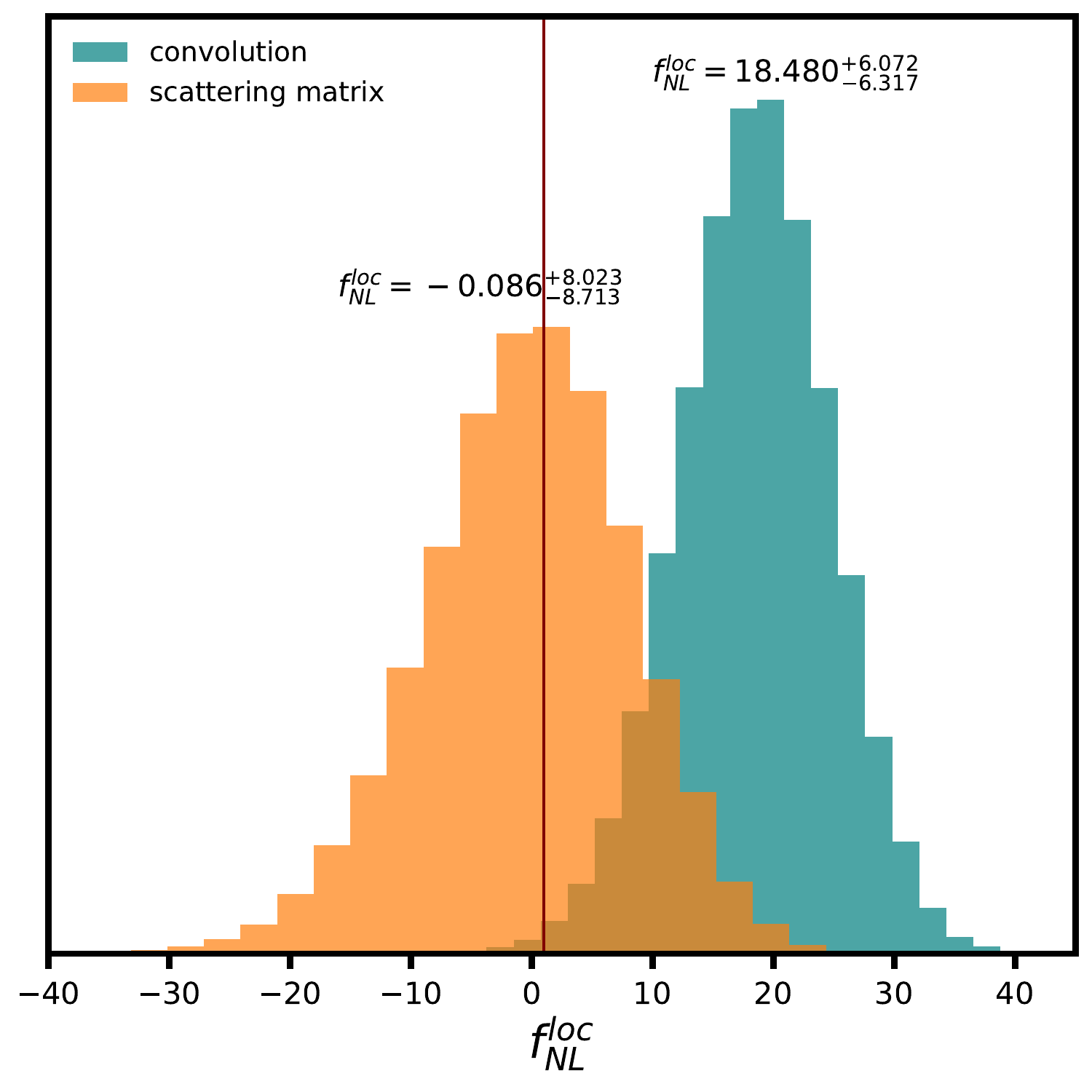}{0.33\textwidth}{}
          \fig{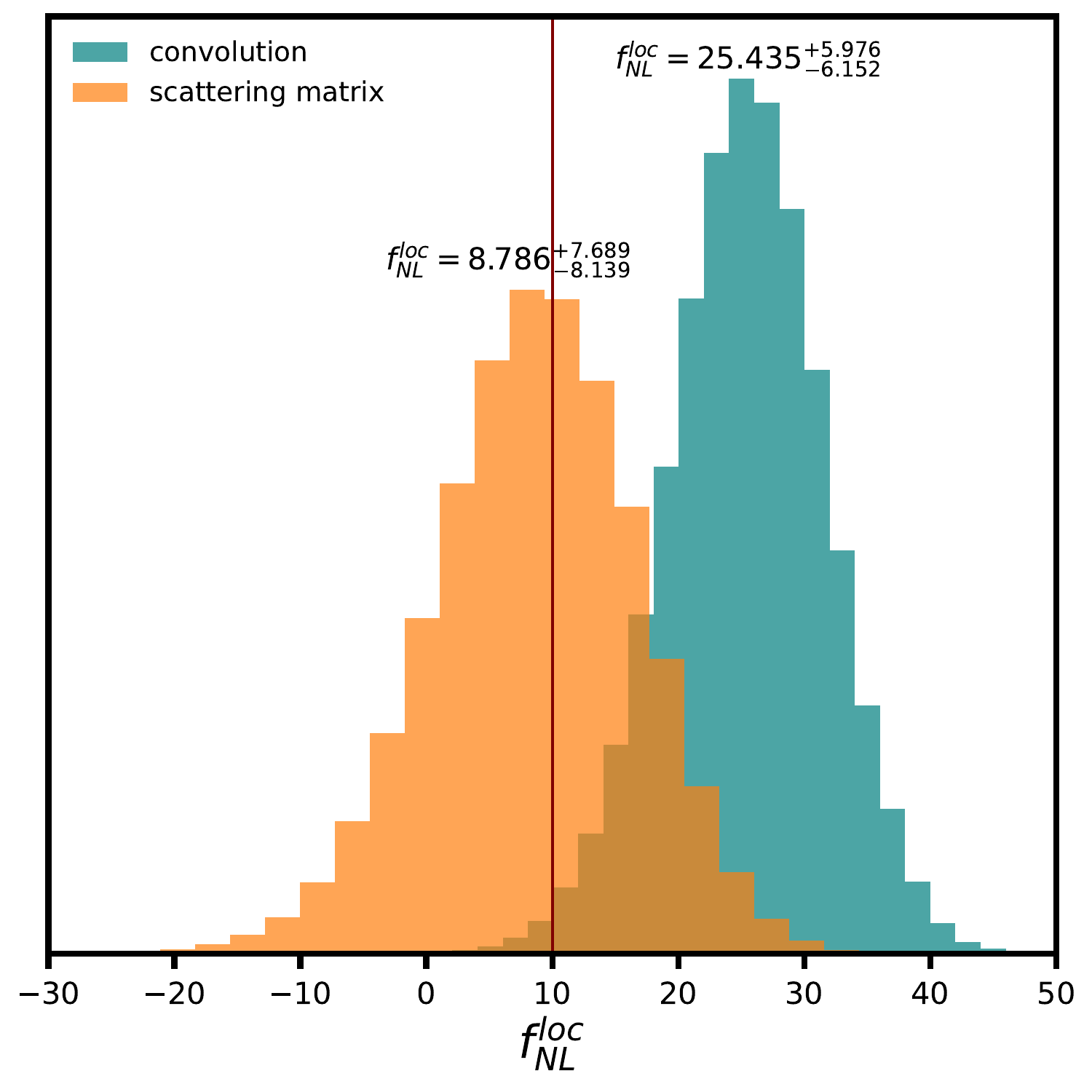}{0.33\textwidth}{}
          \fig{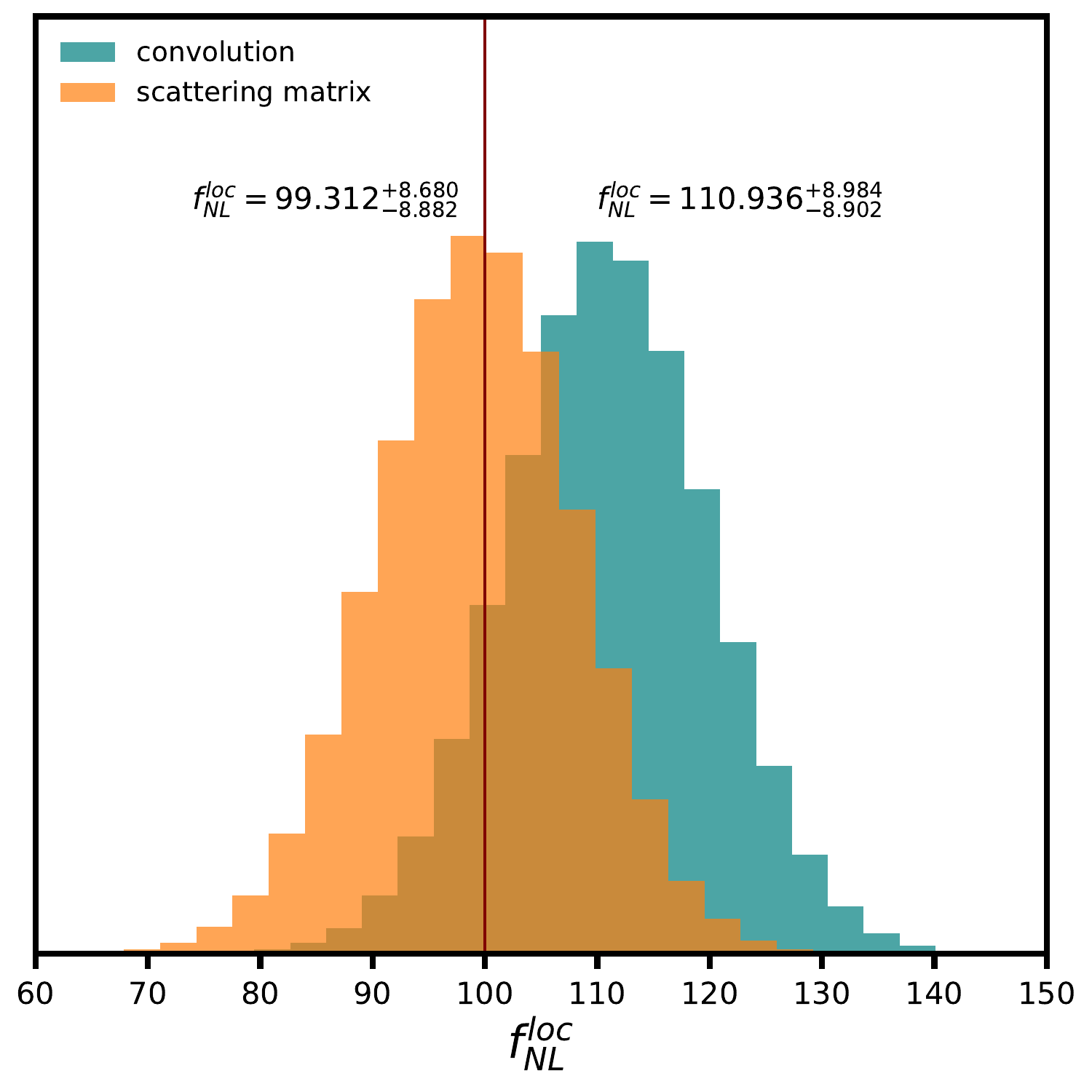}{0.33\textwidth}{}}
\gridline{\fig{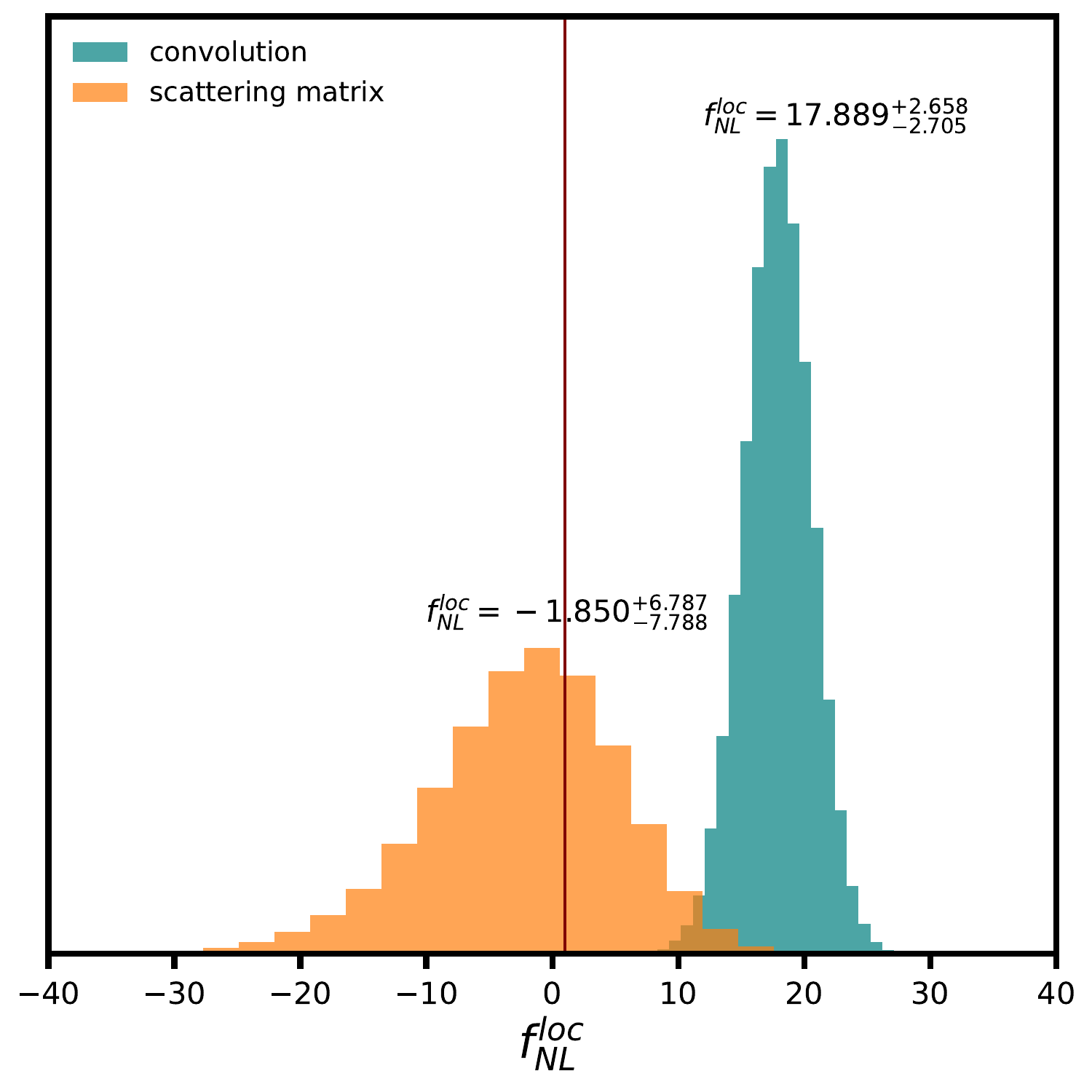}{0.33\textwidth}{(a) $f_{\text{NL}}^{\text{loc,true}} = 1$}
          \fig{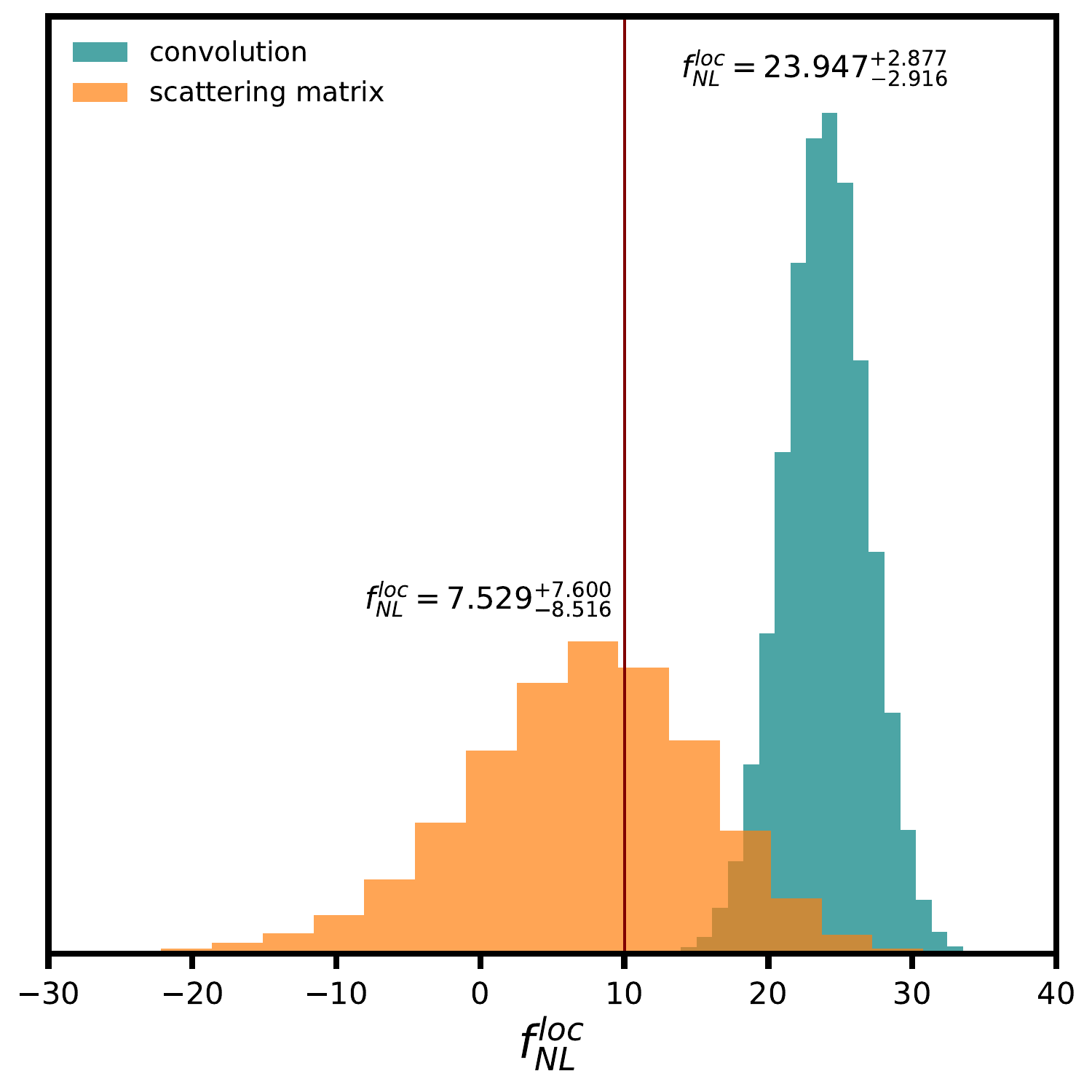}{0.33\textwidth}{(b) $f_{\text{NL}}^{\text{loc,true}} = 10$}
          \fig{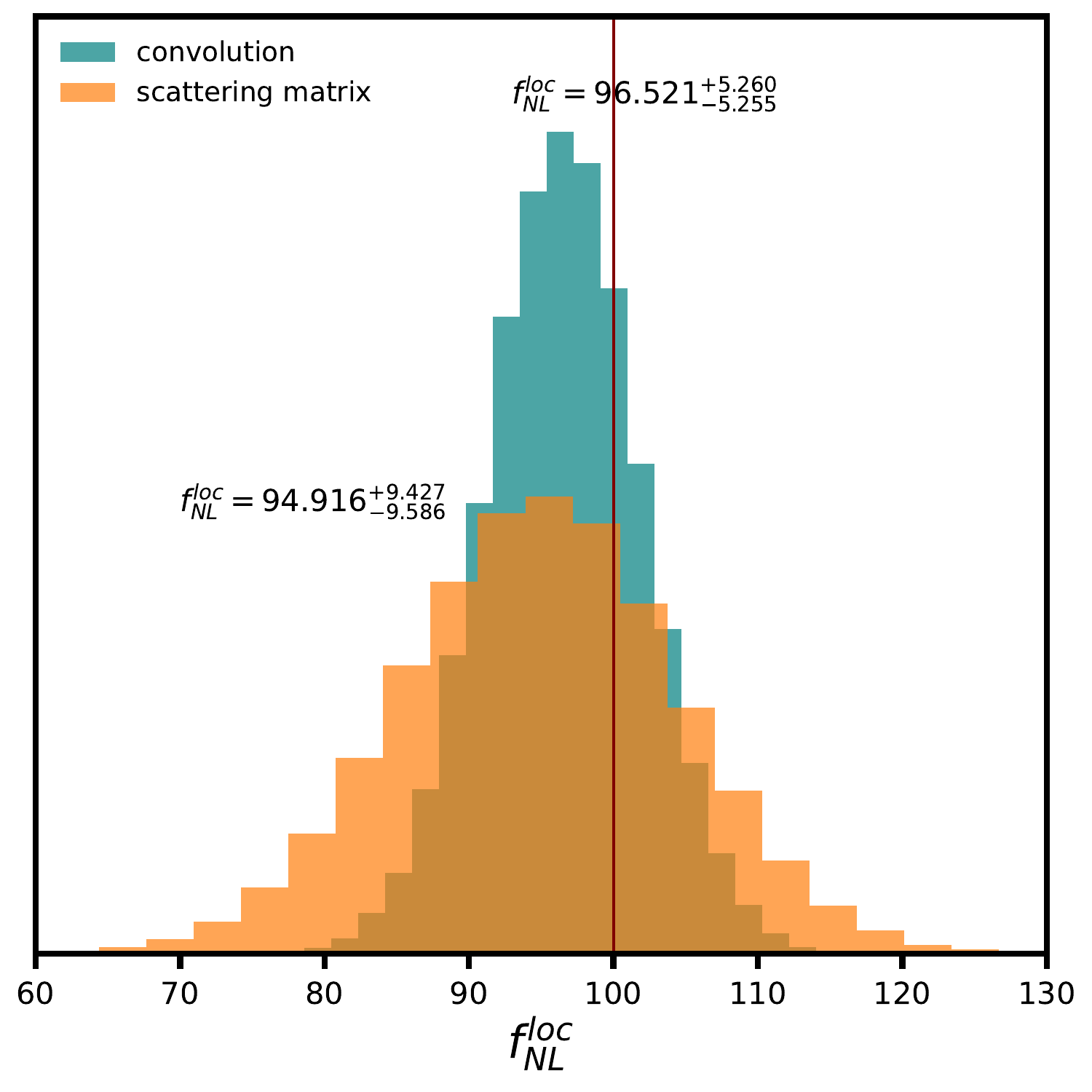}{0.33\textwidth}{(c) $f_{\text{NL}}^{\text{loc,true}} = 100$}}
\caption{{The best-fit values of $f_{\text{NL}}^{\text{loc}}$ parameter estimated from the average power spectra of $500$ realisations after adding photo-$z$ errors. The upper and lower panels correspond to Case-I and Case-II simulations (see main text for description of Case-I and Case-II). The vertical red line marks the true value of $f_{\text{NL}}^{\text{loc}}$ parameter used in simulations. The green histograms are the posteriors obtained following the convolution method to account for photo-$z$ errors. The orange histograms are the posteriors obtained from the scattering matrix approach.}}
\label{fig:comp_fnl_with_wout_corr}
\end{figure}

\begin{figure}[ht!]
\gridline{\fig{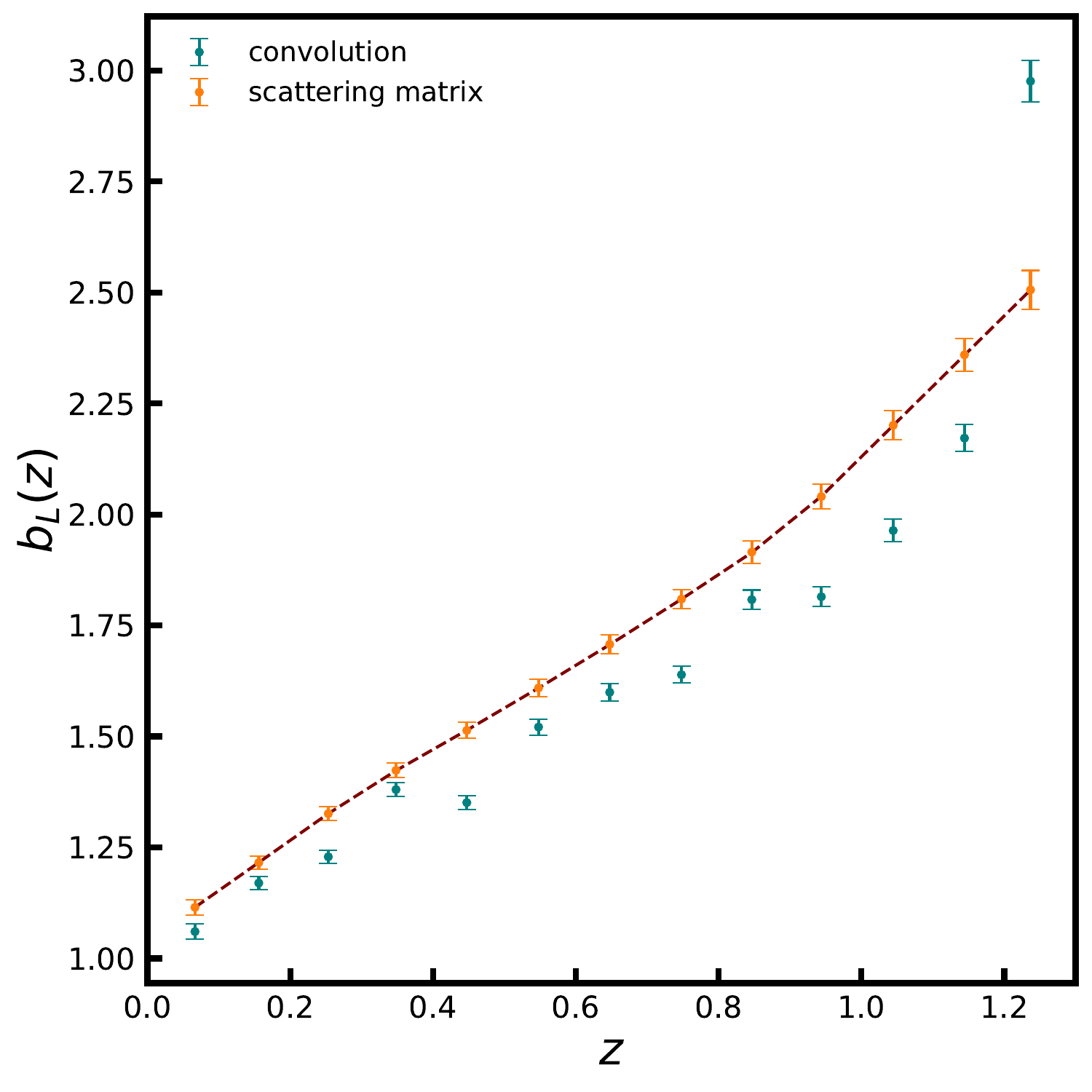}{0.33\textwidth}{}
          \fig{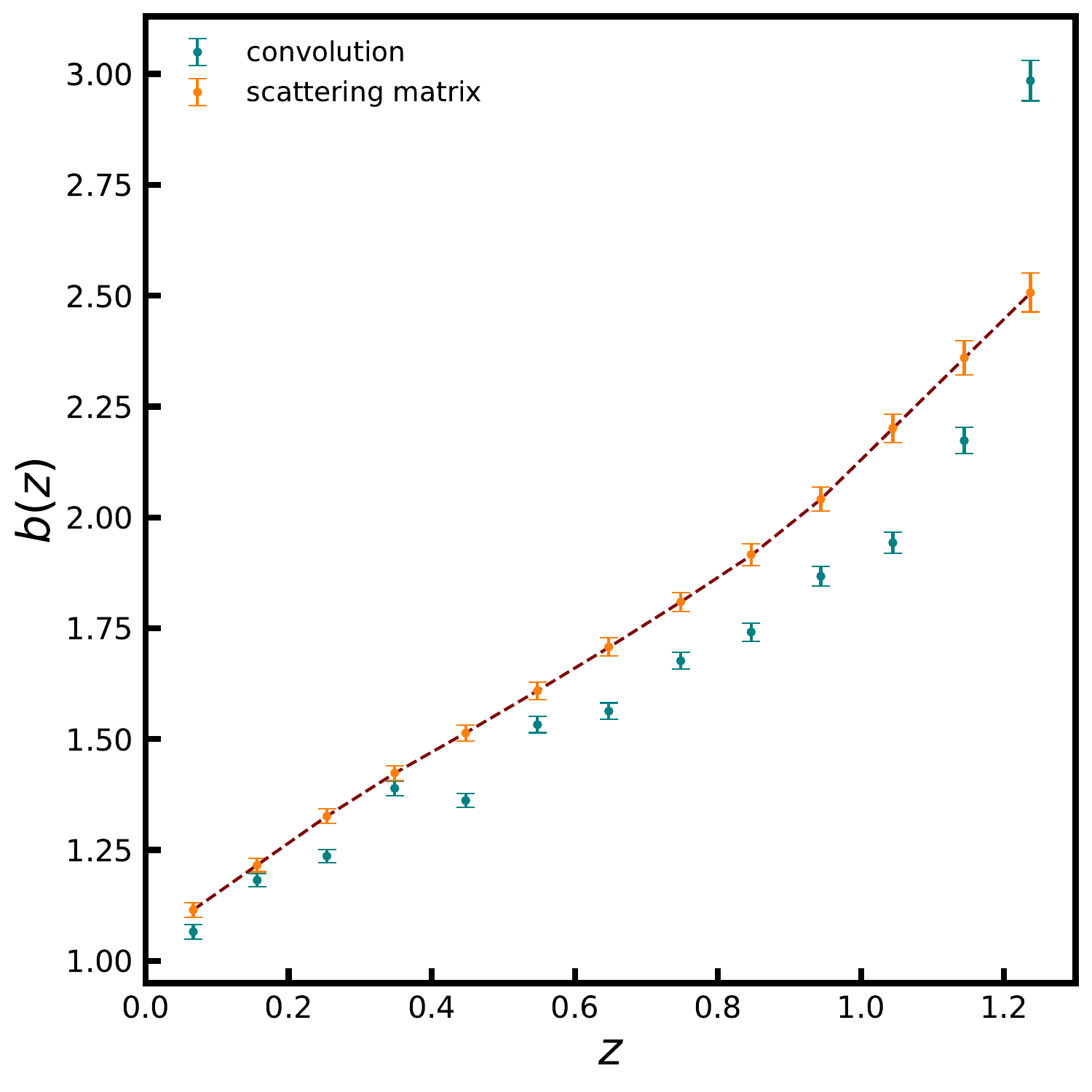}{0.33\textwidth}{}
          \fig{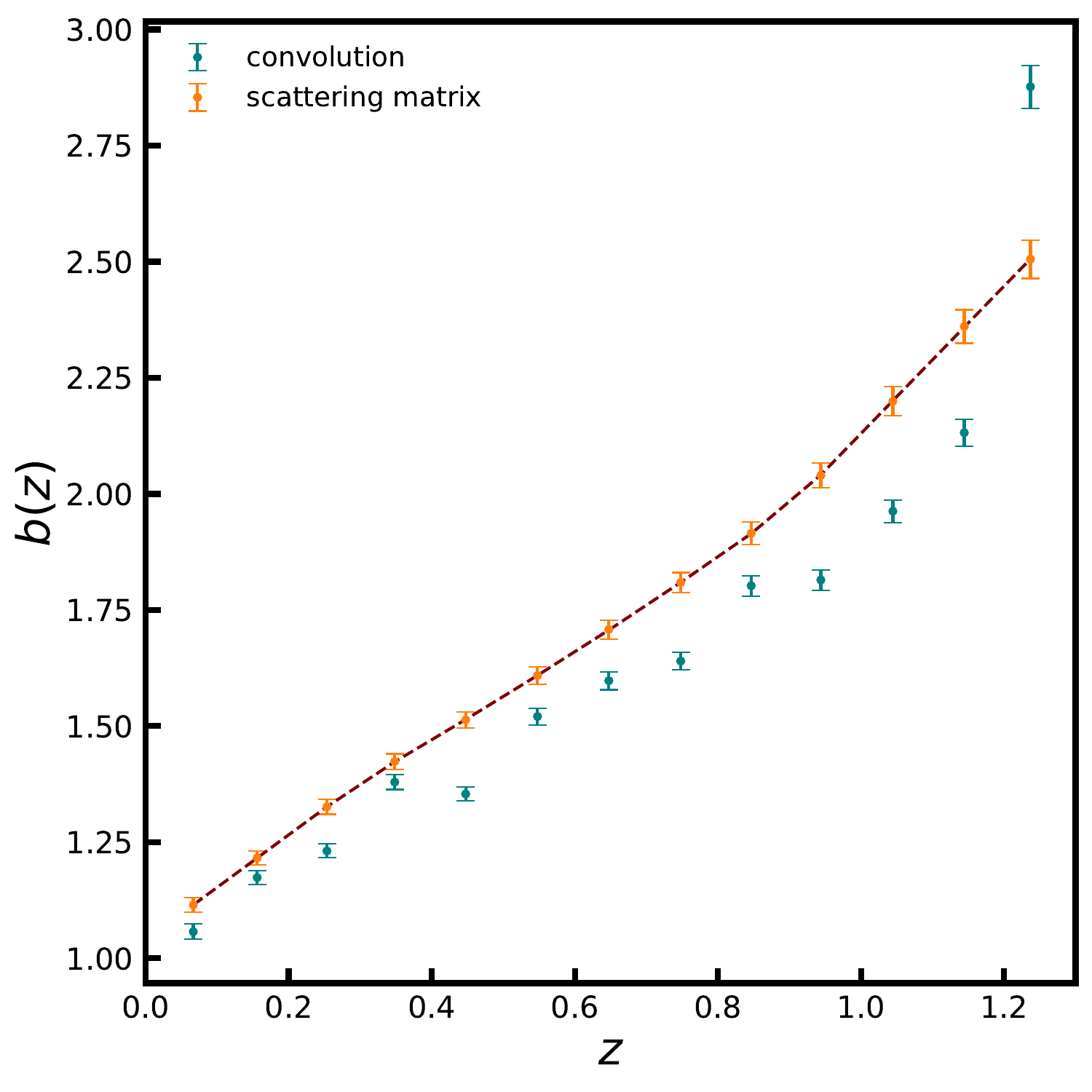}{0.33\textwidth}{}}
\gridline{\fig{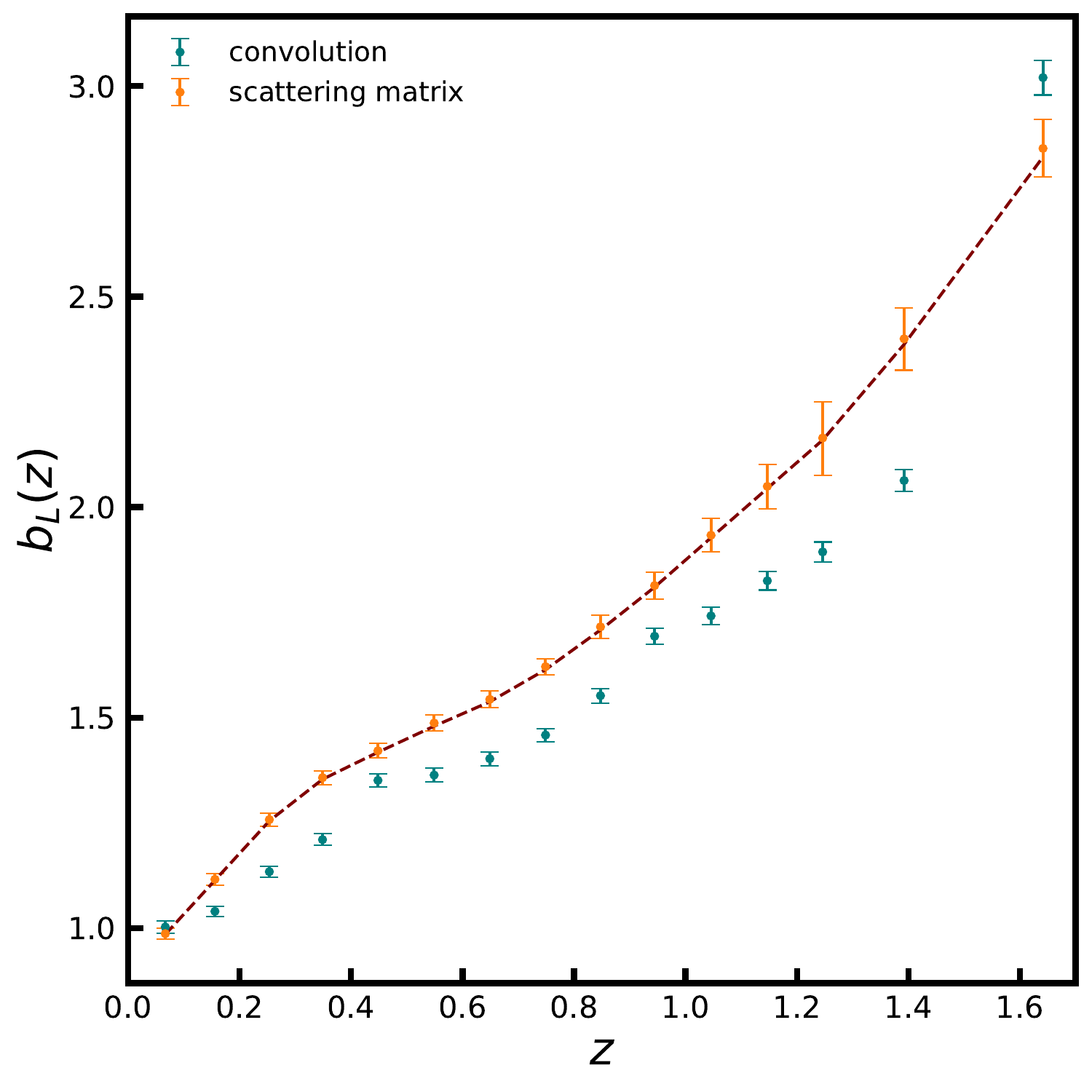}{0.33\textwidth}{(a) $f_{\text{NL}}^{\text{loc,true}} = 1$}
          \fig{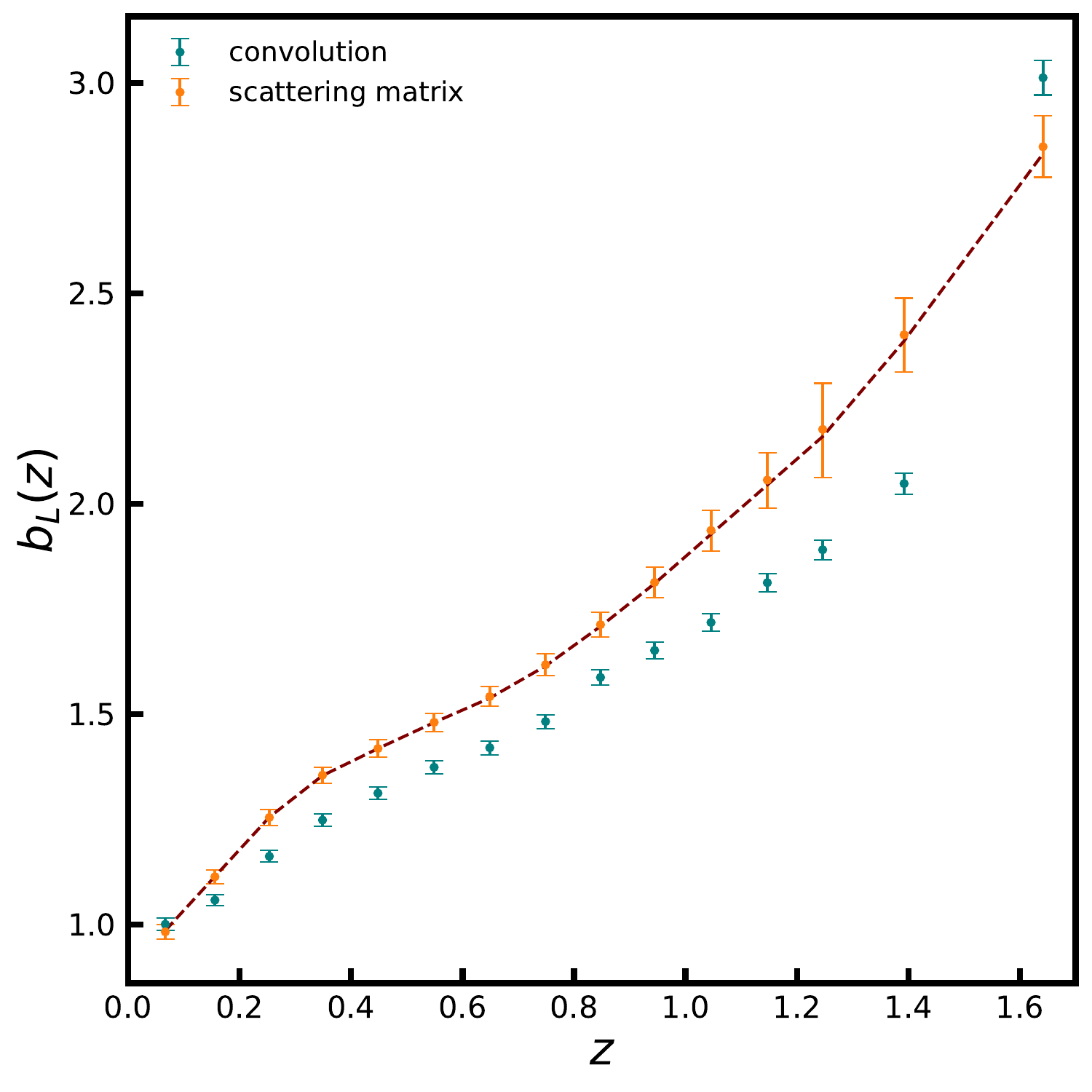}{0.33\textwidth}{(b) $f_{\text{NL}}^{\text{loc,true}} = 10$}
          \fig{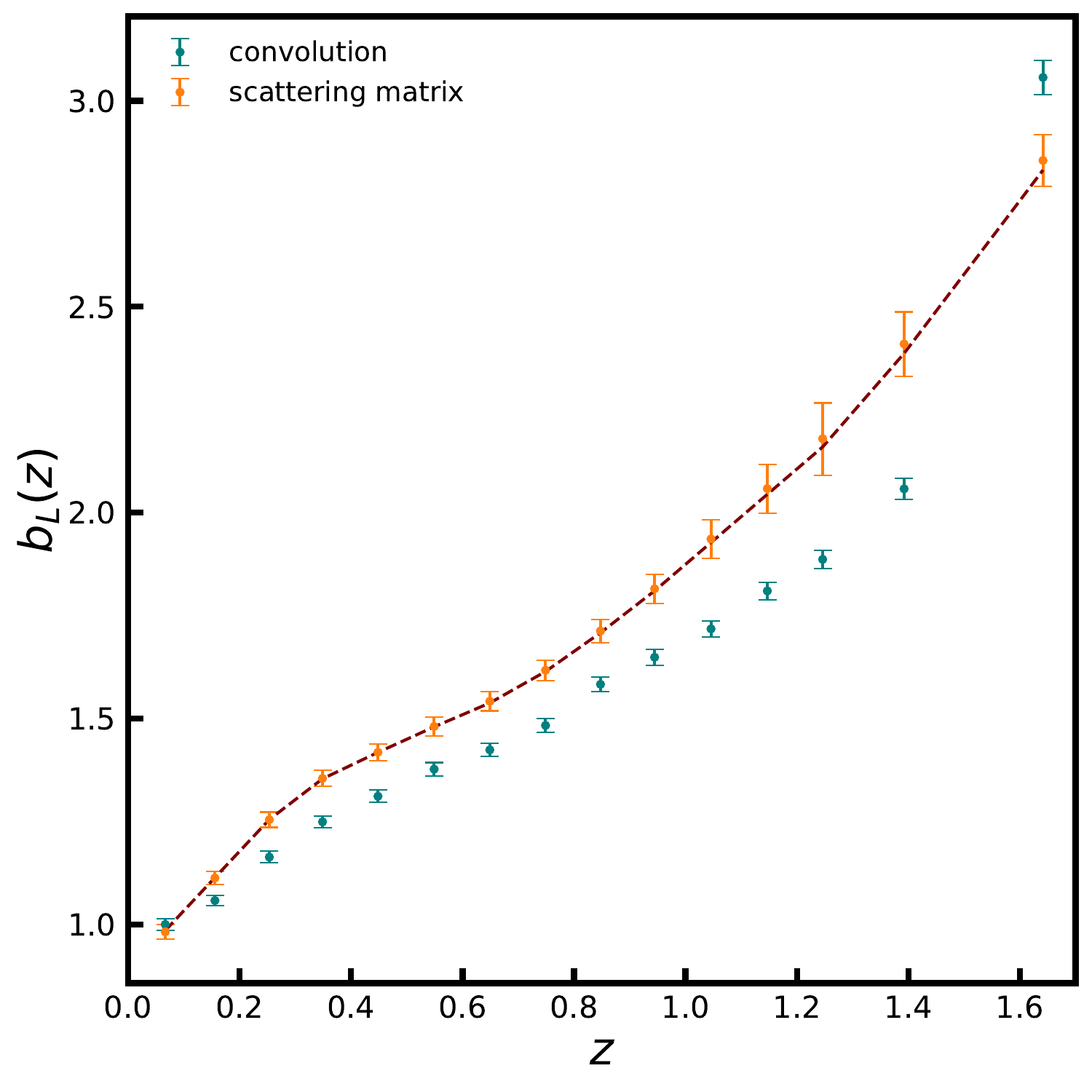}{0.33\textwidth}{(c) $f_{\text{NL}}^{\text{loc,true}} = 100$}}
\caption{{The galaxy halo bias evolution estimated from the average power spectra of $500$ realisations after adding photo-$z$ errors. The upper and lower panels correspond to Case-I and Case-II simulations (see main text for description of Case-I and Case-II). The dashed red line marks the fiducial evolution of $b_{L}(z)$. The green circles are best-fit values of galaxy bias obtained from the convolution method, while the orange circles are those estimated with the scattering matrix approach.}}
\label{fig:comp_gal_bias_with_wout_corr}
\end{figure}

{The posteriors distributions for $f_{\text{NL}}^{\text{loc}}$ after incorporating photo-$z$ errors are shown in Figure \ref{fig:comp_fnl_with_wout_corr} (green histograms). The upper and lower panels correspond to Case-I and Case-II, respectively. The vertical red lines mark the fiducial values of $f_{\text{NL}}^{\text{loc}}$ used in the simulations. Similarly, constraints on the galaxy bias after including photo-$z$ errors are presented in Figure \ref{fig:comp_gal_bias_with_wout_corr} with green circles. The dashed red line represents the fiducial evolution of galaxy bias assumed in our simulations. In Case-I, we observe $1$--$3,\sigma$ shifts in the estimated $f_{\text{NL}}^{\text{loc}}$ values, even after accounting for photo-$z$ errors as outlined in Section \ref{sec:propagate_photoz_errors}. The deviations in galaxy bias are more pronounced, reaching up to $\sim 9\sigma$ from the fiducial model. We note that this trend in galaxy bias is consistent across all three fiducial values of $f_{\text{NL}}^{\text{loc}}$, as the photo-$z$ error prescription remains unchanged.} 

{In Case-II, the discrepancy in $f_{\text{NL}}^{\text{loc}}$ increases, reaching up to $\sim 6\sigma$ for $f_{\text{NL}}^{\text{loc,true}} = 1$ and $10$. Interestingly, the estimate of $f_{\text{NL}}^{\text{loc}}$ for $f_{\text{NL}}^{\text{loc,true}} = 100$ remains consistent with the fiducial value within $1\sigma$ uncertainty. However, the galaxy halo bias exhibits even stronger deviations, up to $\sim 12\sigma$, across all values of $f_{\text{NL}}^{\text{loc,true}}$. This demonstrates that including low redshift-accuracy bins to boost the number density of sources can still yield biased constraints on both $f_{\text{NL}}^{\text{loc}}$ and the galaxy bias. The increased galaxy number count is effectively counterbalanced by the greater overlap in tomographic redshift distributions. The complete set of 1D and 2D posterior distributions for Case-I are provided in Figures \ref{fig:full_likelihood_fnl_1}--\ref{fig:full_likelihood_fnl_100}.}

\begin{figure}[ht!]
\gridline{\fig{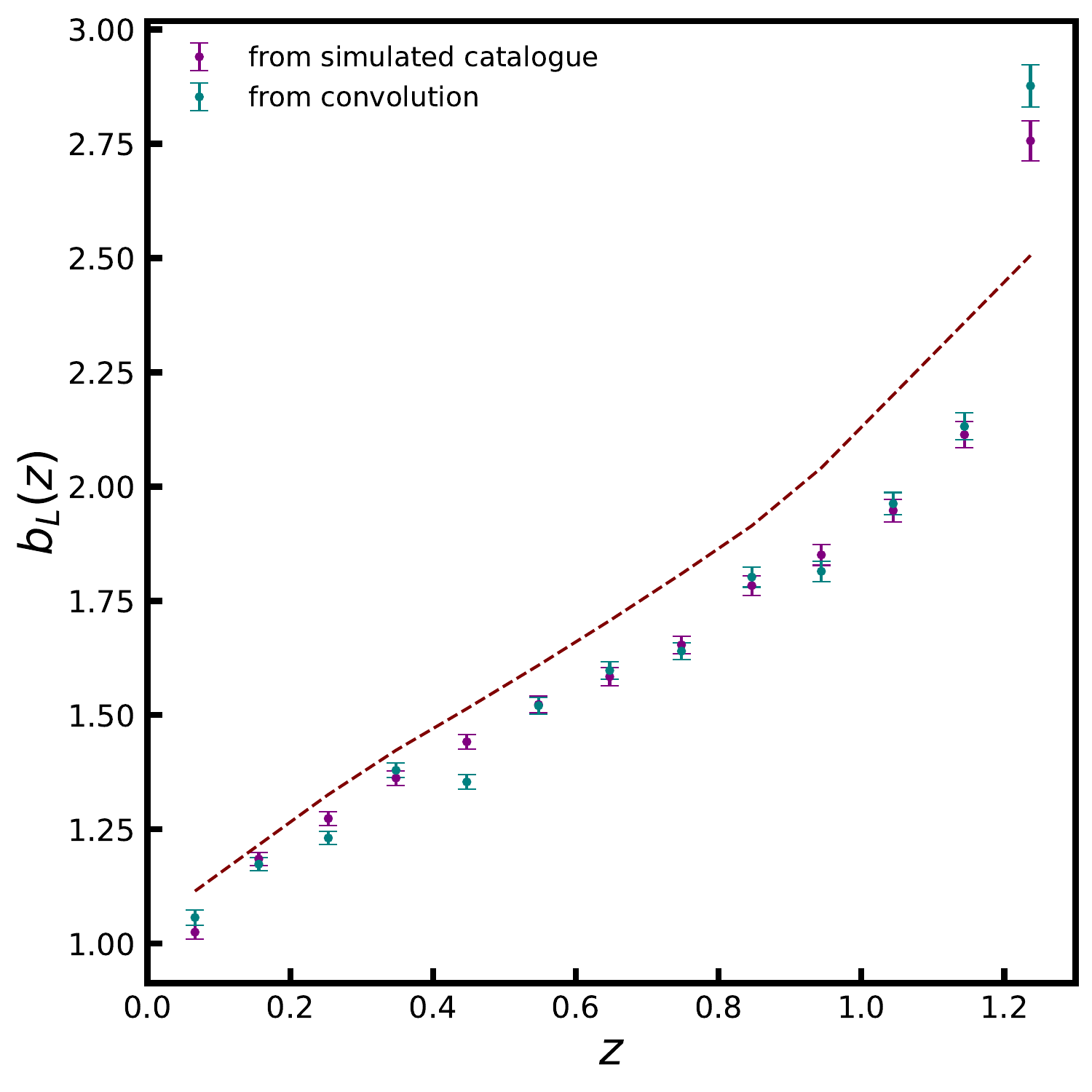}{0.5\textwidth}{}
          \fig{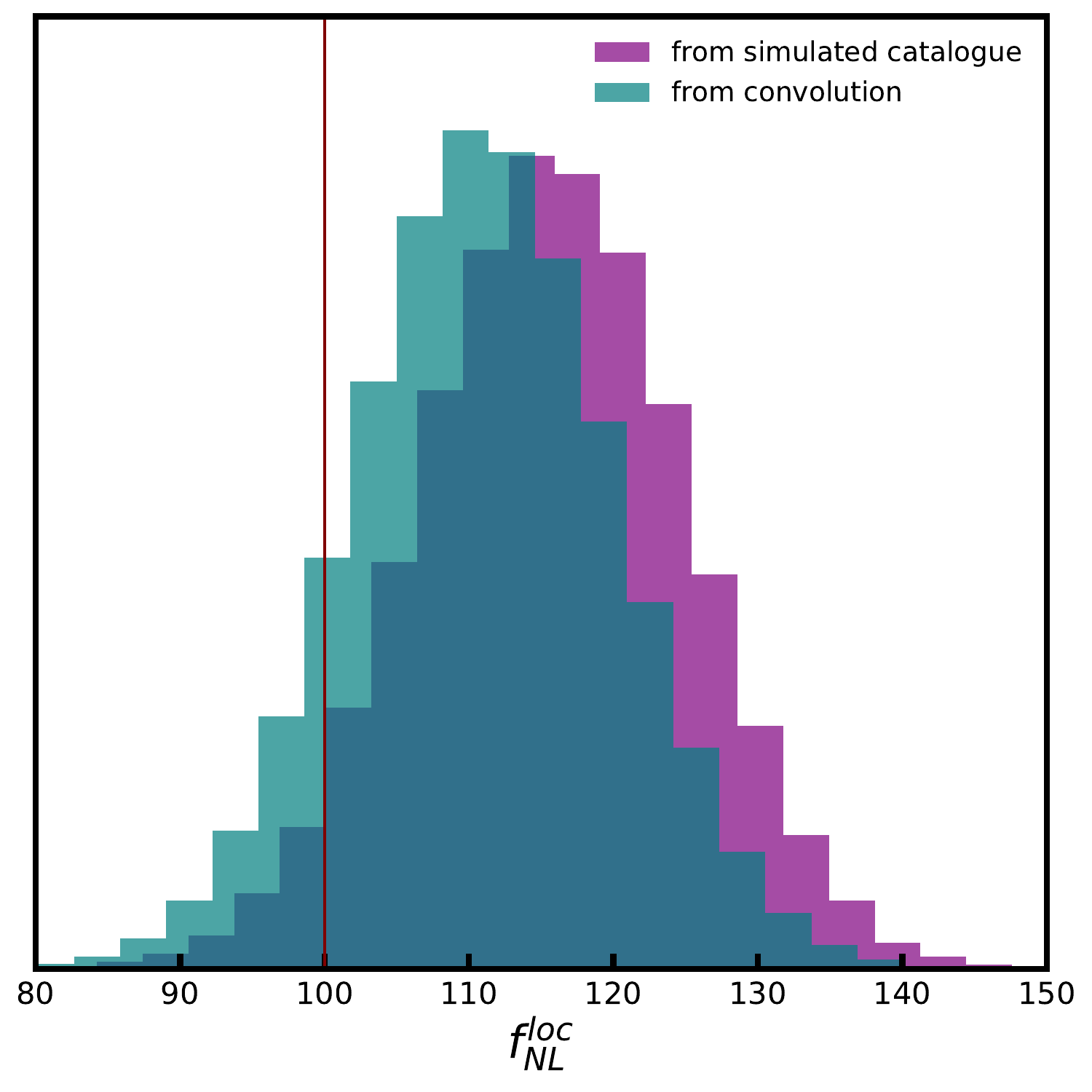}{0.5\textwidth}{}}
\caption{{Comparison of galaxy bias (left panel) in tomographic bins and $f_{\text{NL}}^{\text{loc}}$ (right panel) estimated from redshift distribution computed via the convolution method (green circles) and by tracking galaxies in simulated catalogue (purple circles). The red lines mark the true values of parameters used for simulations.}}
\label{fig:compare_params_from_convolution_and_simulated_catalogue}
\end{figure}

{We emphasize that cross-power spectra between redshift bins were not utilized to refine the tomographic redshift distributions in Eq.~\ref{eq:true_dist_conv}. To test the robustness of our parameter estimates against uncertainties in the redshift distribution, we replaced the convolved distribution $\frac{\mathrm{d}N^{i}(z)}{\mathrm{d}z}$ (Section \ref{sec:propagate_photoz_errors}) with one derived directly from the simulated catalogue by tracking individual galaxies. This substitution allows us to replace the estimated true redshift distribution with its ground-truth counterpart from the simulation. Figure~\ref{fig:compare_params_from_convolution_and_simulated_catalogue} compares the resulting parameter estimates from the convolution method and by tracking galaxies in simulated catalogue, for $f_{\text{NL}}^{\text{loc,true}}=100$ in Case-I. Notably, using the catalogue-derived redshift distribution does not improve parameter constraints.} This indicates that the differences in the parameter estimates do not stem from any biased estimates of $\frac{\mathrm{d}N^{i}(z)}{\mathrm{d}z}$. These differences are instead due to systematics present in the data which we refer to as ``redshift bin mismatch''.

\begin{figure}[ht!]
\plotone{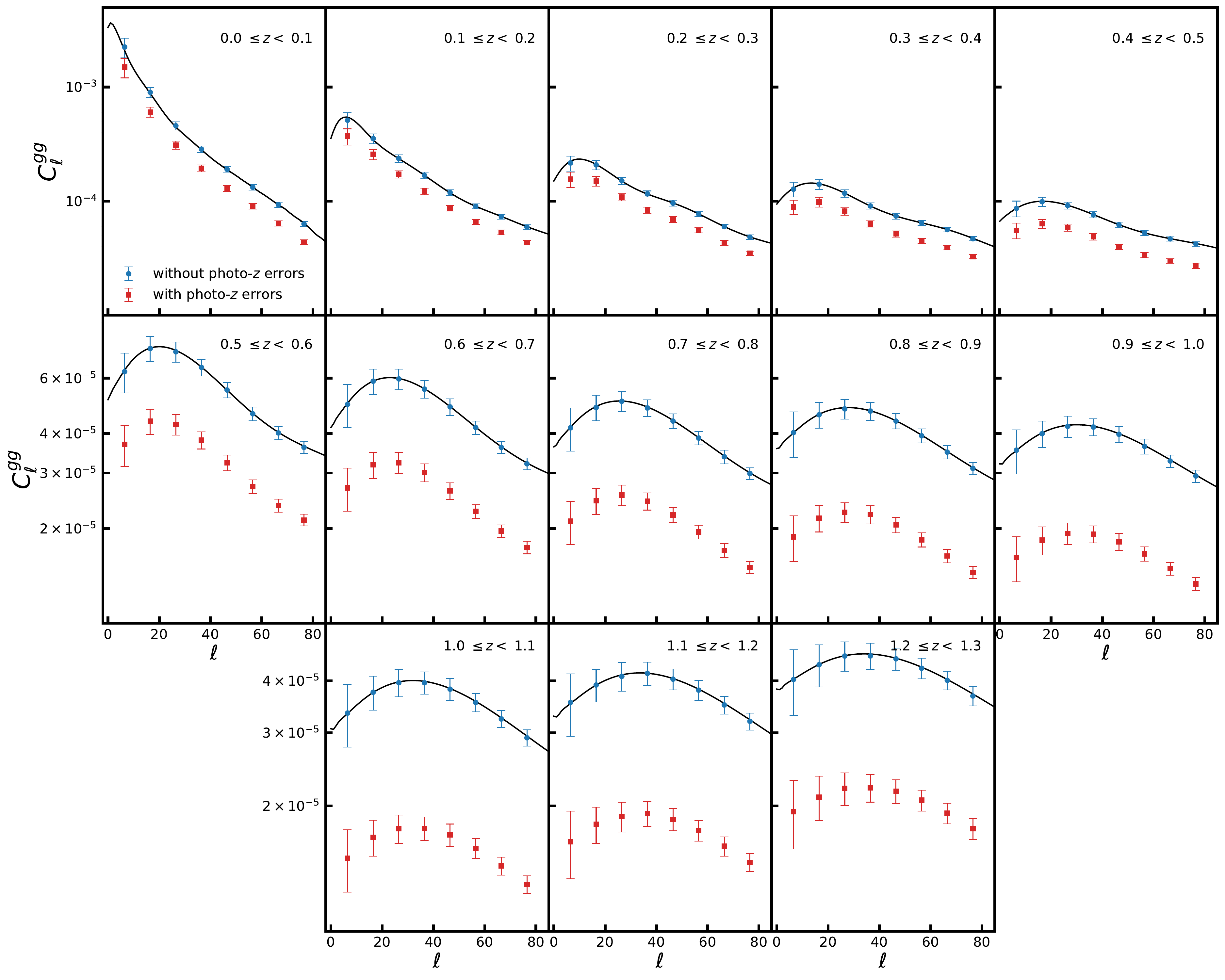}
\caption{The galaxy angular power spectrum measured from $500$ simulations. The black line represents the underlying true power spectrum. The blue circles are the power spectra estimated before adding photo-$z$ errors. The red squares show the power spectra after adding photo-$z$ errors. The error bars on the data points are standard errors computed from the diagonal of the sample covariance matrices (Eq. \ref{eq:sample_covariance}).}
\label{fig:comp_cl_with_wout_photoz_errors}
\end{figure}

The observed parameter offsets are due to the fact that photo-$z$ errors cause galaxies to be classified in wrong redshift bins.
{The diffusion of galaxies across tomographic bins leads to substantial discrepancies from the true underlying power spectrum across all scales, as illustrated in Figure \ref{fig:comp_cl_with_wout_photoz_errors}. We performed a detailed investigation of the impact of redshift bin mismatch on the $\sigma_{8}$ parameter, along with mitigation strategy, in C24. In the following section, we provide a brief overview of the mitigation approach developed to achieve unbiased measurements of the $f_{\text{NL}}^{\text{loc}}$ parameter.}

\subsection{Correction for redshift bin mismatch}\label{sec:scattering_matrix_approach}
\begin{figure}[ht!]
\gridline{\fig{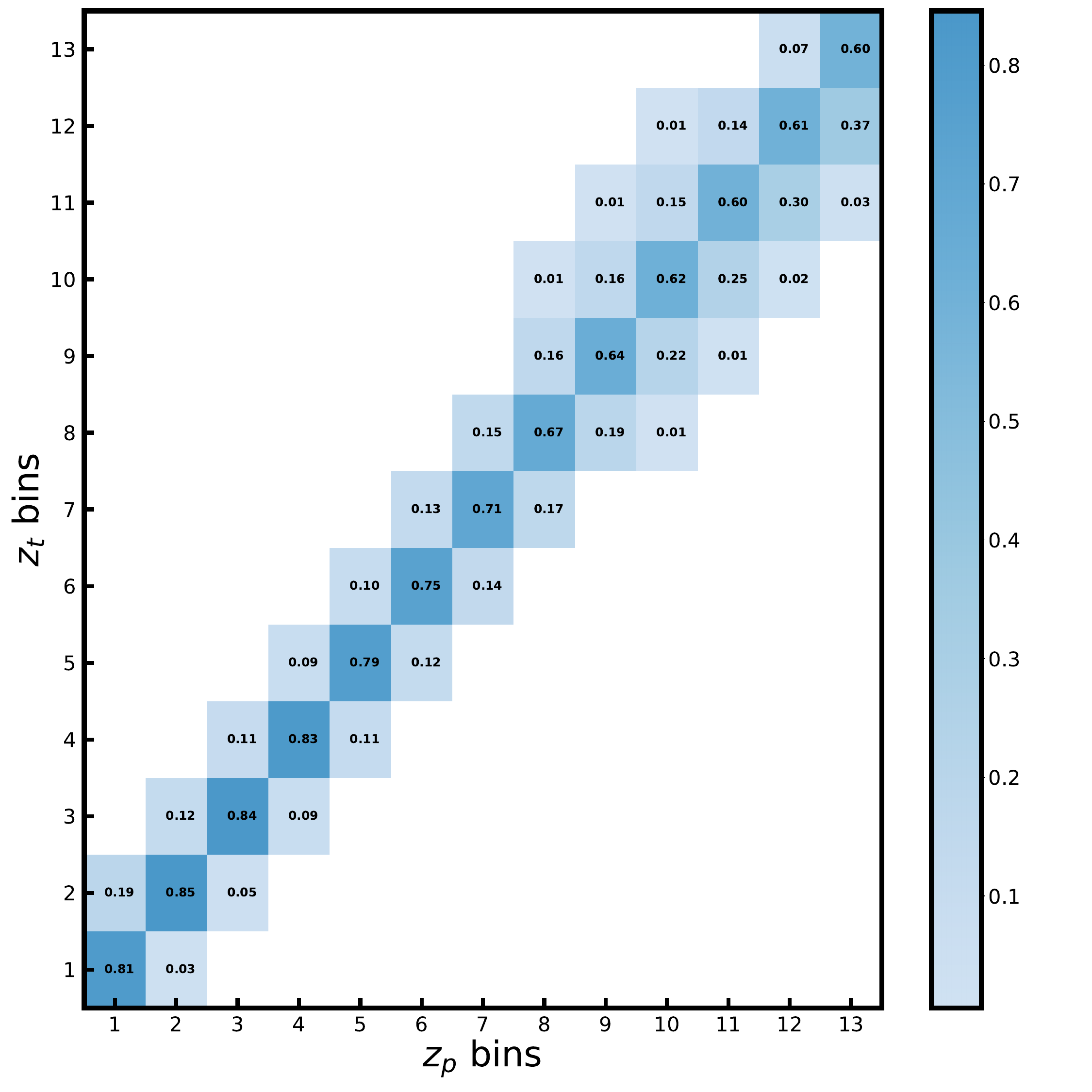}{0.5\textwidth}{(a) Case-I}
          \fig{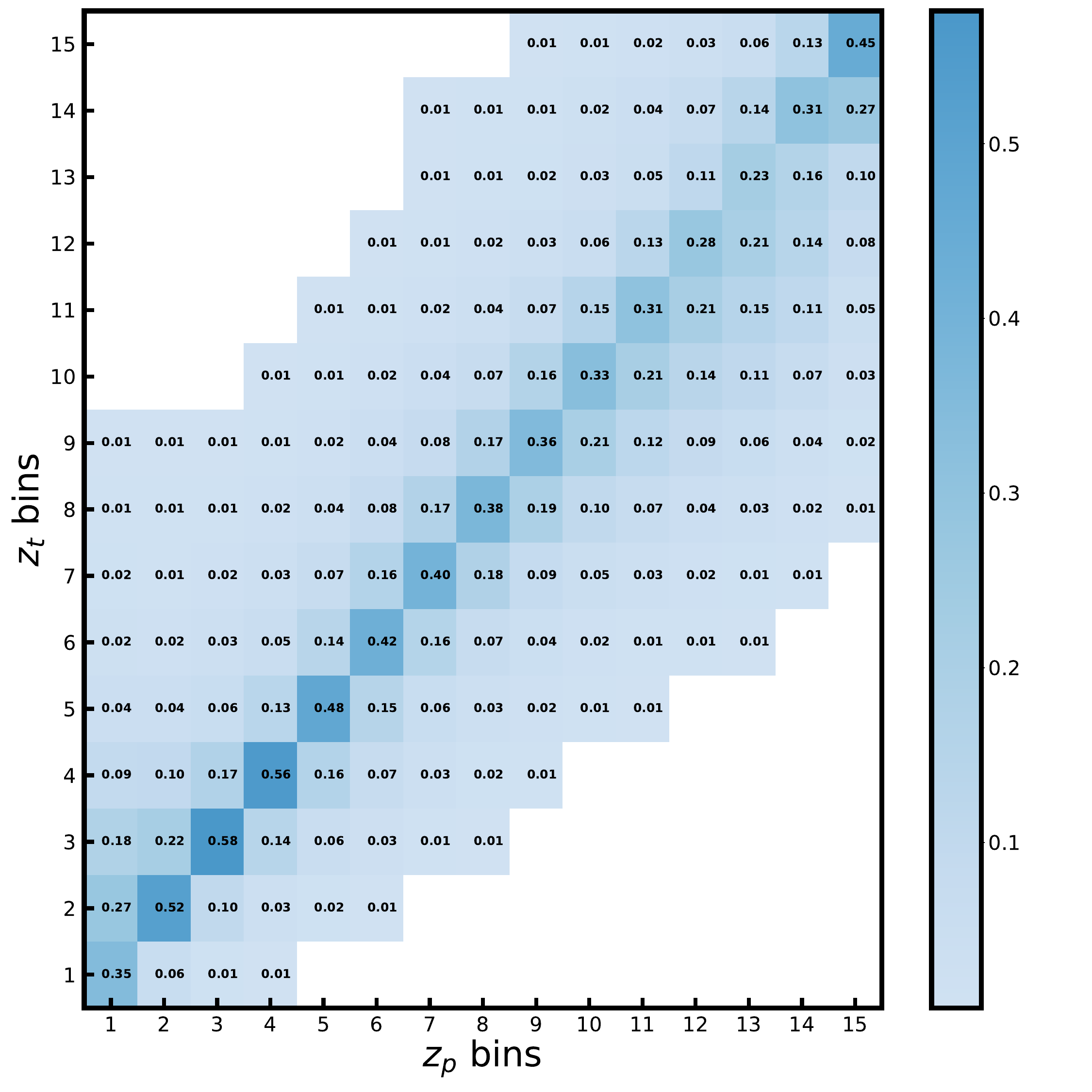}{0.5\textwidth}{(b) Case-II}}
\caption{The average scattering matrix estimated from $500$ simulations using Eq. \ref{eq:scattering_matrix_element_equation}.}
\label{fig:mean_scatterring_matrix_fnl_1}
\end{figure}

As mentioned earlier, the parameters estimated in a tomographic angular correlation analyses will be biased due to galaxies ending up in the wrong redshift bin owing to their photo-$z$ errors. The redshift bin mismatch couples the power spectra computed from true redshifts $C^{gg,\text{tr}}(\ell)$ to that from photometric redshifts $C^{gg,\text{ph}}(\ell)$ \citep{2010MNRAS.405..359Z}:
\begin{equation}
    C_{ij}^{gg,\text{ph}}(\ell) = \sum\limits_{x,y}P_{xi}P_{yj}C_{xy}^{gg,\text{tr}}(\ell),
    \label{eq:scattering_relation_gg}
\end{equation}
where $i,j,x,y$ denote tomographic bins and $P_{ij}$ represents the fraction of galaxies moving from redshift bin $i\to j$ due to photo-$z$ errors. We can re-write Eq. \ref{eq:scattering_relation_gg} in the matrix form as:
\begin{equation}
    C^{gg,\text{ph}} = P^\top C^{gg,\text{tr}}P
    \label{eq:scattering_relation_gg_matrix}
\end{equation}
where $P$ is now called the scattering matrix. {In C24, we followed the deconvolution method to directly compute the coefficients $P_{ij}$ from the ratio}
\begin{equation}
    P_{ij} = \int\limits_{z_{\text{min}}^{j}}^{z_{\text{min}}^{j+1}}\mathrm{d}z \frac{\mathrm{d}N^{i}}{\mathrm{d}z}\bigg/ {\int\limits_{z_{\text{min}}^{j}}^{z_{\text{min}}^{j+1}}\mathrm{d}z_{p} \frac{\mathrm{d}N}{\mathrm{d}z_{p}}},
    \label{eq:scattering_matrix_element_equation}
\end{equation}
where $\frac{\mathrm{d}N}{\mathrm{d}z_{p}}$ is the photometric redshift distribution of the galaxies, and $z_{min}^{j}$ is the lower limit of the $j$th redshift bin. {Although straight-forward to use, the deconvolution method requires proper regularisation. In absence of any generalised penalty function, the deconvolution method become challenging for not-so smooth redshift distributions. In this work, we compute the scattering matrix coefficients from the observed photometric redshift distribution and error distributions through a convolution approach. Our new method surpasses the one used in C24 in terms of accuracy and feasibility to compute the true redshift distribution. We refer the readers to C24 for a more detailed explanation on the scattering matrix formalism. The average scattering matrix computed from $500$ simulations for the fiducial value $f_{\text{NL}}^{\text{loc,true}} = 1$ is presented in Figure \ref{fig:mean_scatterring_matrix_fnl_1}. The extent of galaxies scattering across redshift bins can be directly compared between Case-I and Case-II. As expected, the inclusion of low redshift accuracy bins in Case-II results in a higher degree of redshift bin mismatch.}

We present the results of parameters estimation with the scattering matrix formalism in Figure \ref{fig:comp_fnl_with_wout_corr} with orange histograms and Figure \ref{fig:comp_gal_bias_with_wout_corr} with orange circles. {Notably, the scattering matrix formalism provides unbiased estimates of $f_{\text{NL}}^{\text{loc}}$ and accurately recovers the true evolution of galaxy bias for both Case-I and Case-II. It is important to highlight that the power spectra $C^{gg,\text{ph}}$ becomes a linear combination of $C^{gg,\text{tr}}$, weighted quadratically by the matrix elements $P_{ij}$. This mixing of power from different true redshift bins may not be fully captured by a precise estimate of the true redshift distribution alone. Consequently, this misrepresentation can lead to systematic offsets in the derived parameter distribution.}

\subsection{{Impact of redshift binwidth}}

\begin{figure}[ht!]
\gridline{\fig{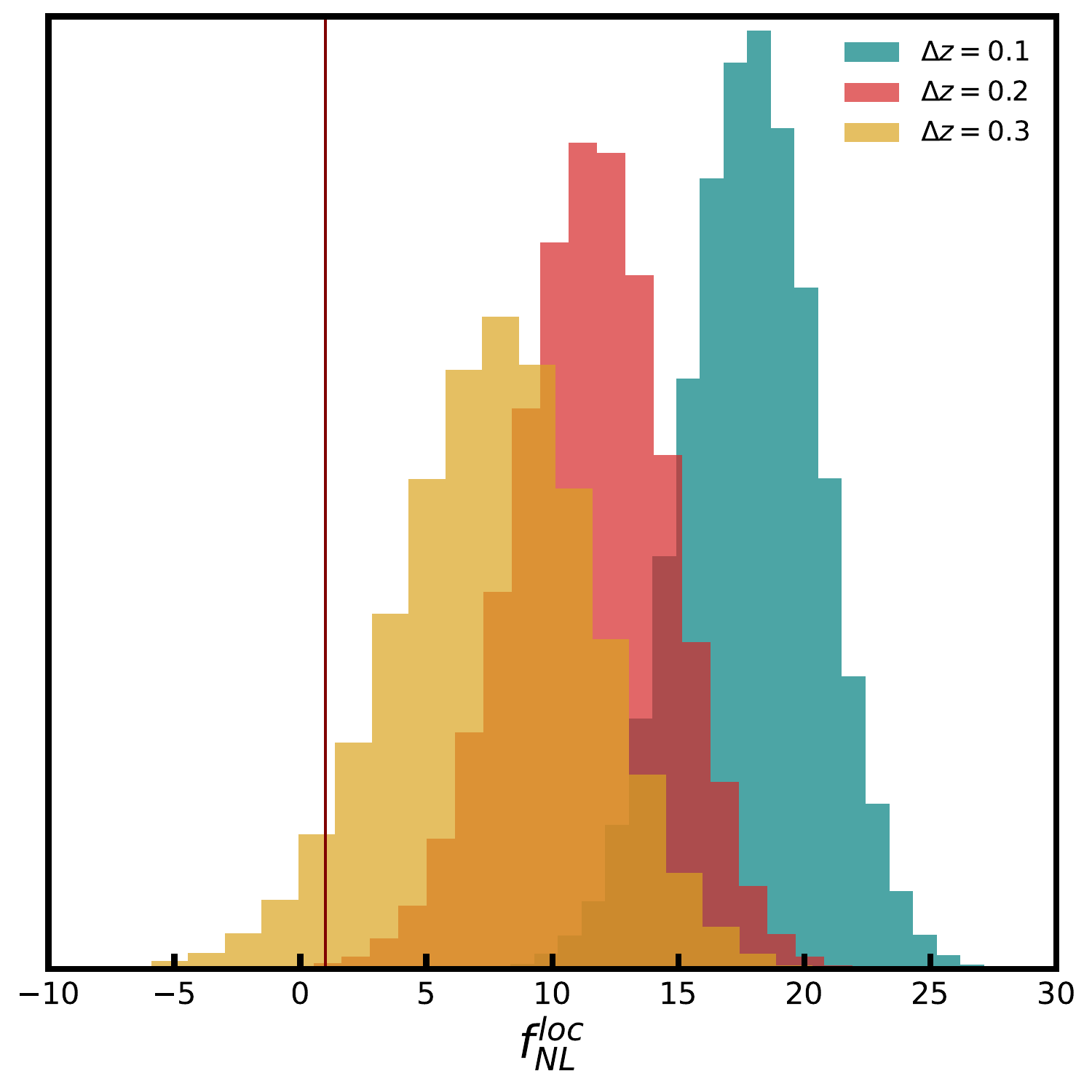}{0.5\textwidth}{(a) convolution}
          \fig{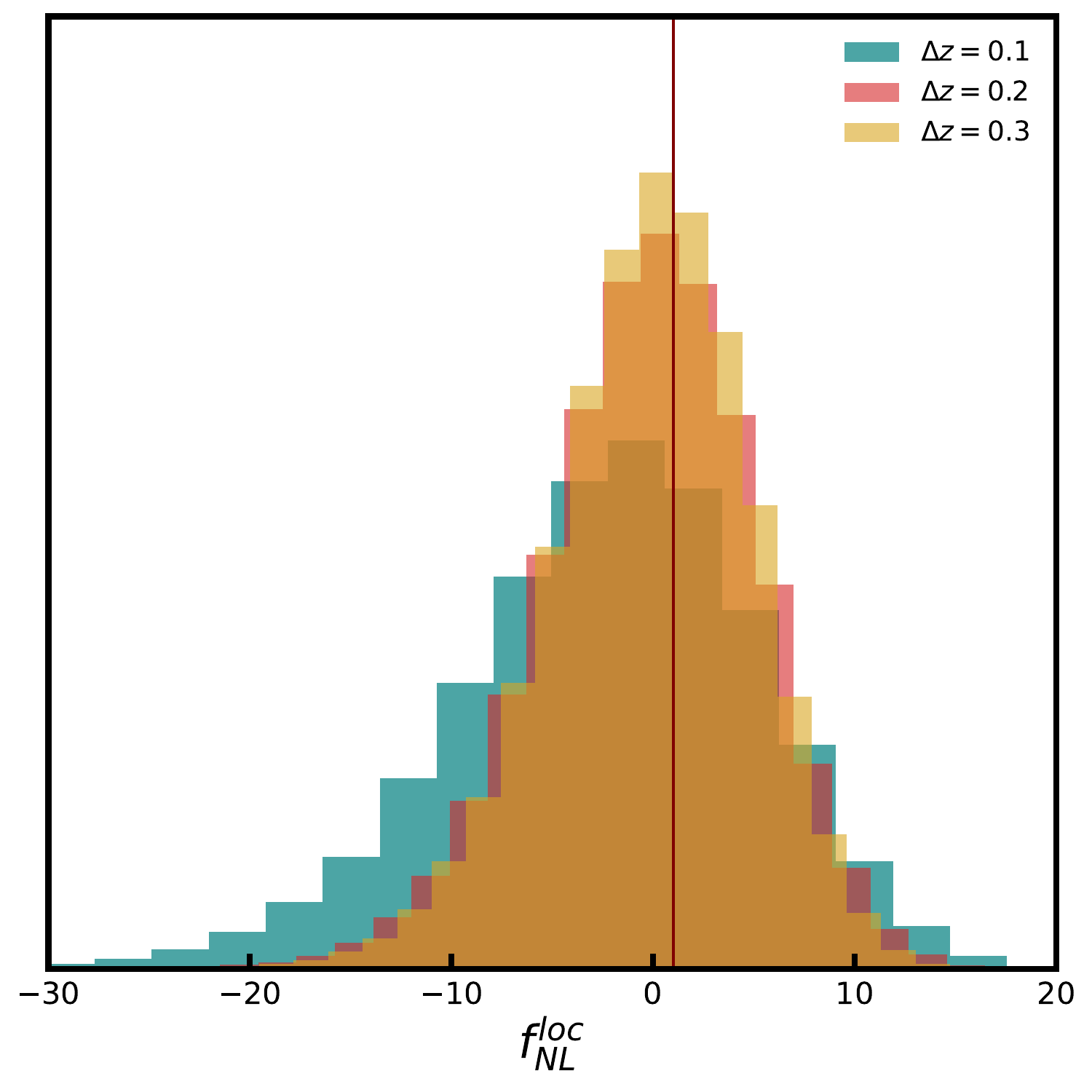}{0.5\textwidth}{(b) scattering matrix}}
\caption{{Effect of the width of the tomographic bins on estimation of $f_{\text{NL}}^{\text{loc}}$ parameter, without $(a)$ and with $(b)$ the scattering matrix correction for redshift bin mismatch. The green, red and yellow histograms show the posteriors of $f_{\text{NL}}^{\text{loc}}$ parameter obtained with tomographic measurements made with redshift bin size $\Delta z=0.1, 0.2 \text{ and } 0.3$, respectively. The vertical line marks the fiducial value of $f_{\text{NL}}^{\text{loc,true}}=1$ used in these simulations.}}
\label{fig:comp_fnl_for_different_redshift_binwidths}
\end{figure}

{To further investigate the scattering of galaxies, we assess how the width of tomographic bins influences estimates of $f_{\text{NL}}^{\text{loc}}$. To achieve this, we take the simulation setup for fiducial $f_{\text{NL}}^{\text{loc,true}}=1$ in Case-II and create 500 mocks each for galaxies divided into 8 and 5 tomographic bins with $\Delta z=0.2 \text{ and } 0.3$, respectively. Figure \ref{fig:comp_fnl_for_different_redshift_binwidths} presents the comparison for $f_{\text{NL}}^{\text{loc}}$ with different redshift binwidths, computed with the convolution method and our scattering matrix formalism. We find that increasing the size of redshift bins reduces the bias on $f_{\text{NL}}^{\text{loc}}$ when using the convolution method. It is an expected outcome because broader tomographic bins dilutes the scatter of galaxies across redshift. However, our scattering matrix formalism fully accounts for redshift bin mismatch, yielding unbiased estimates irrespective of bin width. While larger bins may mitigate the mismatch to some extent, this comes at the cost of reduced sensitivity to redshift evolution of other parameters such as the galaxy bias or $\sigma_{8}$. In contrast, our scattering matrix approach preserves this sensitivity as well as remains robust across varying redshift binning schemes and photo-$z$ accuracies. Therefore, we advocate for the adoption of the scattering matrix formalism in future analyses of tomographic angular clustering for cosmological parameter estimation.}

\section{Summary}\label{sec:conclusions}
The differentiation between single and multi-field inflationary scenarios is hinged upon the measurements of local primordial non-Gaussianity. A tight constraint of $\sigma(f_{\text{NL}}^{\text{loc}})\sim 1$, required to understand the dynamics of inflaton field, will be possible with the combinations of future CMB experiments and large-scale structure surveys. The next-generation of photometric surveys such as LSST and SPHEREx will play a pivotal role in measuring $f_{\text{NL}}^{\text{loc}}$ by the means of the tomographic angular clustering measurements. However, as demonstrated in \cite{2024A&A...687A.150S} and \cite{2024A&A...690A.338S}, tomographic measurements suffer from misclassification of galaxies into redshift bins due to photometric redshift errors. {In this work, we forecast the constraints on $f_{\text{NL}}^{\text{loc}}$ in the presence of redshift bin mismatch caused by these errors.}

We prepared $500$ log-normal galaxy density simulations using \texttt{GLASS}, ensuring the physical properties were consistent with the specifications of SPHEREx. {We generated photometric redshifts for galaxies assuming Gaussian error distribution for two different sets of simulations. In Case-I, we used the first three SPHEREx redshift accuracy bins and divided the galaxies in 13 redshift bins. In Case-II, we used all five SPHEREx redshift accuracy bins resulting in 15 redshift bins. The true redshift distribution was estimated from the photometric redshift distribution using the convolution method (Section \ref{sec:propagate_photoz_errors}). Constraints on $f_{\text{NL}}^{\text{loc}}$ were then derived from the measured angular power spectra in tomographic bins using the maximum likelihood estimator.}

Due to diffusion of galaxies across redshift bins, the measured galaxy angular power spectra differ significantly than the underlying true angular power spectra (Figure \ref{fig:comp_cl_with_wout_photoz_errors}). {This diffusion results in $1-3\,\sigma$ offsets for $f_{\text{NL}}^{\text{loc}}$ and up to $\sim 9\sigma$ deviations for galaxy linear halo bias for Case-I, while for Case-II the increase to $1-6,\sigma$ for $f_{\text{NL}}^{\text{loc}}$ and deviations of up to $\sim 12\sigma$ for galaxy halo bias. We observed similar offsets in $f_{\text{NL}}^{\text{loc}}$ and galaxy halo bias when redshift distributions were computed directly from the simulated catalogue. In Section \ref{sec:scattering_matrix_approach}, we demonstrated that our scattering matrix formalism can mitigate the redshift bin mismatch, recovering both $f_{\text{NL}}^{\text{loc}}$ and galaxy bias within $1\sigma$ errors for both Case-I and Case-II.}

We point out that more conventional approaches to estimating the true redshift distribution, such as convolution or deconvolution, are insufficient to address the redshift bin mismatch of galaxies. In tomographic measurements using photometric surveys, bin mismatch can lead to apparent tensions on parameters and biased inferences about cosmological models. We, therefore, propose that the scattering matrix formalism be used for future tomographic studies. This work focused on the impact of photometric redshift errors on $f_{\text{NL}}^{\text{loc}}$ using only the angular power spectrum only. {We plan to explore the impact of other survey systematics, such as catastrophic redshift errors, photometric calibration errors, and joint forecasts using two- and three-point angular correlation, in future studies.}

\begin{acknowledgments}
The authors thank Bomee Lee for stimulating discussions on constraining primordial non-Gaussianity with SPHEREx.
\end{acknowledgments}

%

\vspace{5mm}


\software{\texttt{GLASS} \citep{2023OJAp....6E..11T},  
          \texttt{emcee} \citep{2013PASP..125..306F},
          }



\appendix

\section{Pipeline validation results}\label{appndx:pipeline_validation}

\begin{figure}[ht!]
\gridline{\fig{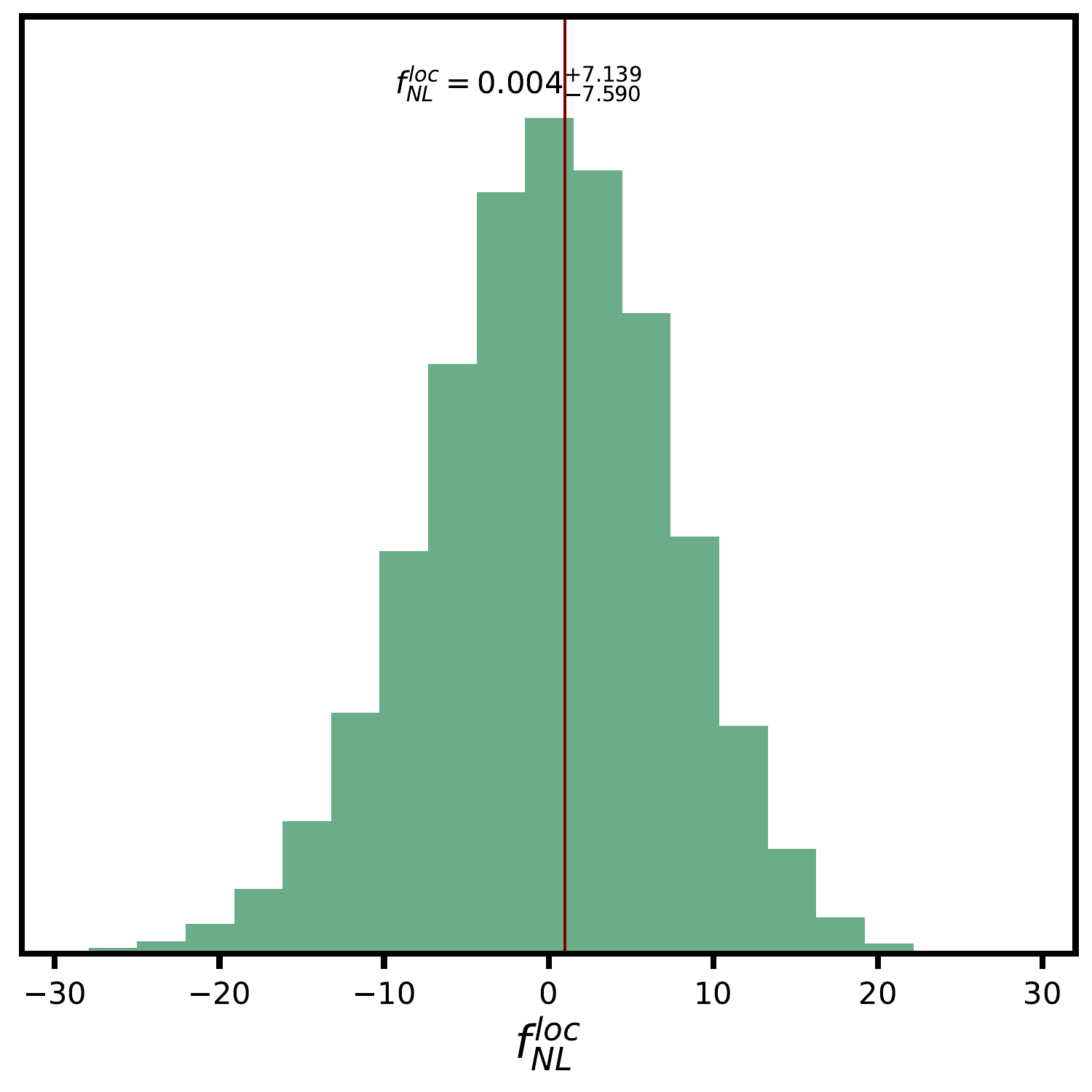}{0.33\textwidth}{(a) $f_{\text{NL}}^{\text{loc,true}} = 1$}
          \fig{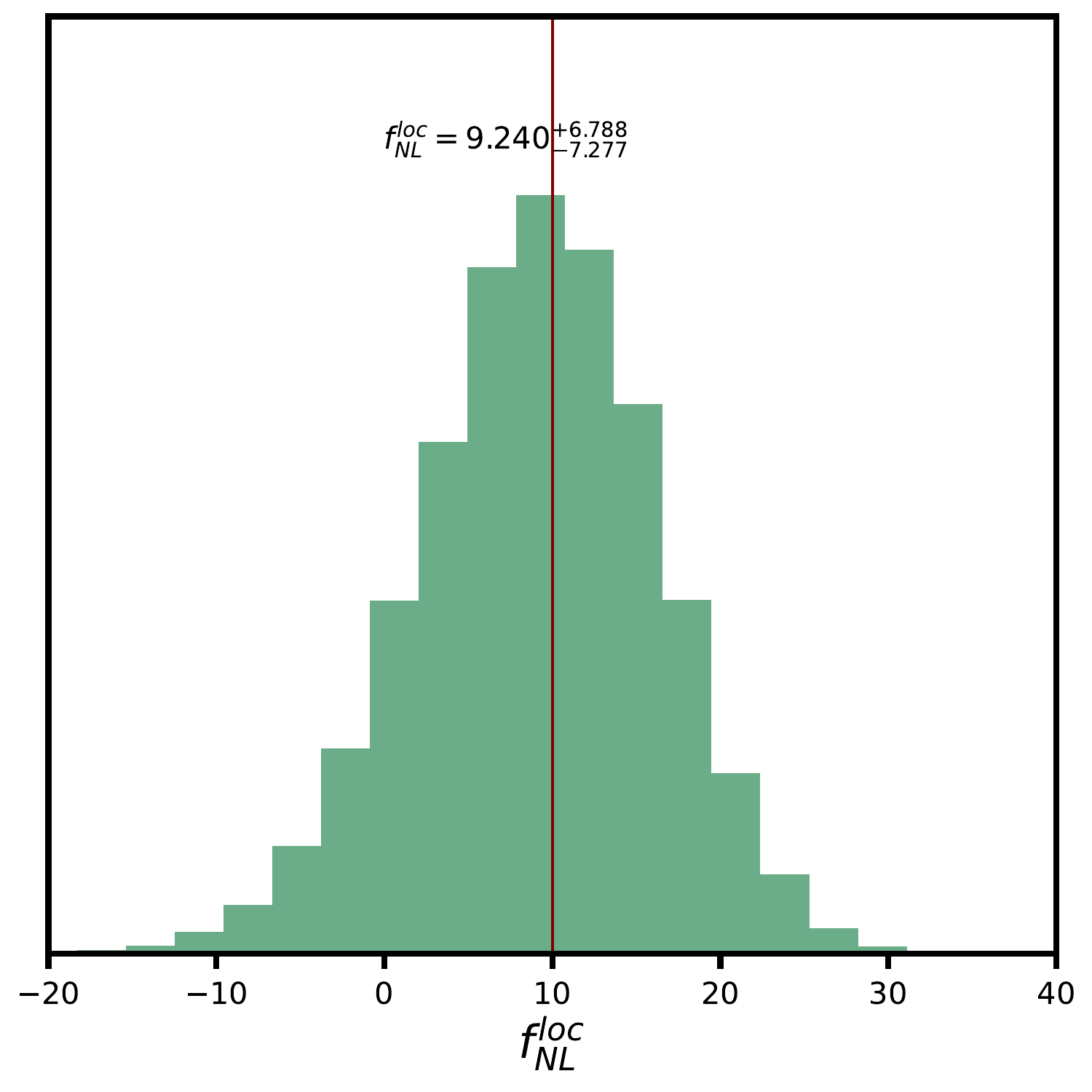}{0.33\textwidth}{(b) $f_{\text{NL}}^{\text{loc,true}} = 10$}
          \fig{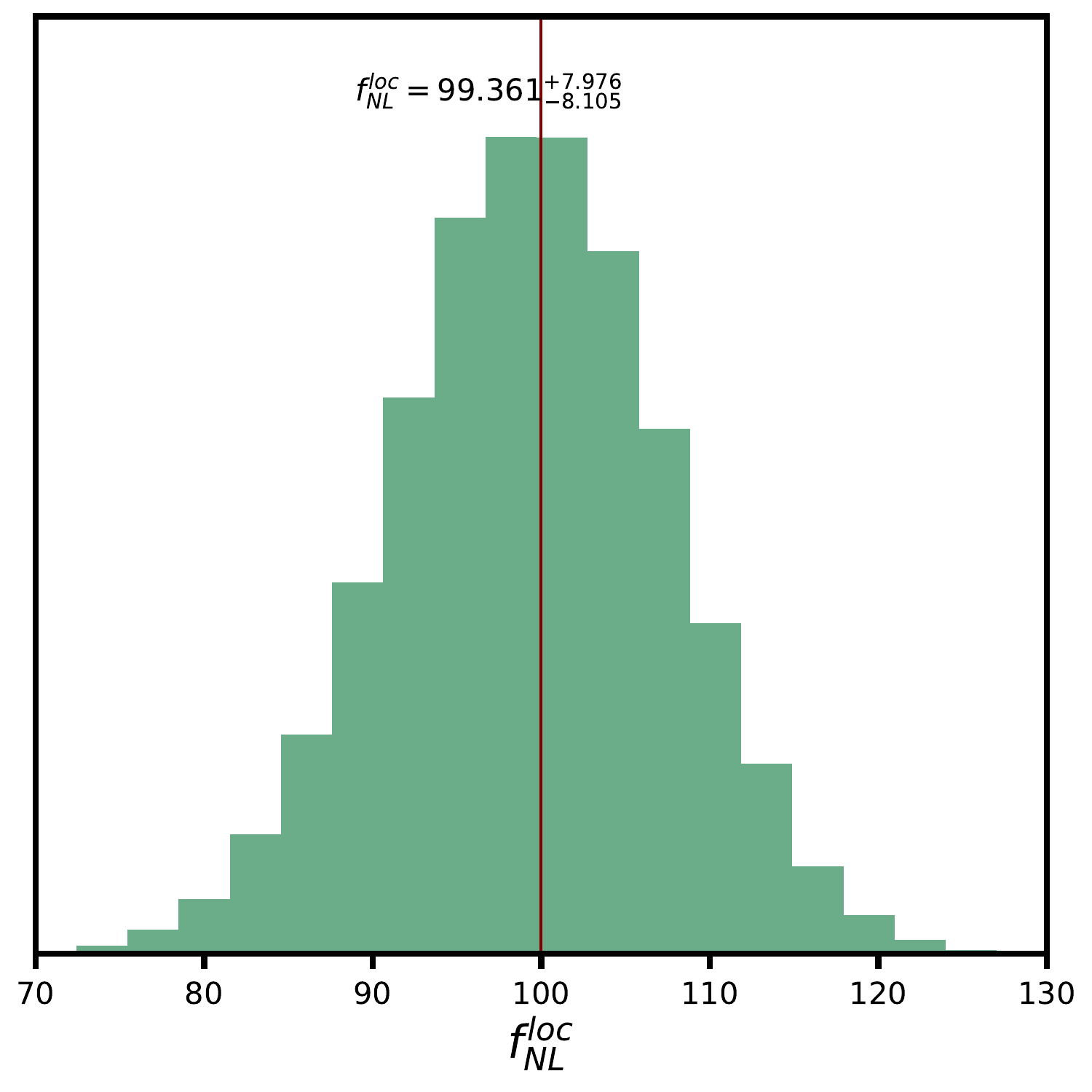}{0.33\textwidth}{(c) $f_{\text{NL}}^{\text{loc,true}} = 100$}}
\caption{The best-fit values of $f_{\text{NL}}^{\text{loc}}$ parameter estimated from the average power spectra of $500$ realisations before adding photo-$z$ errors. The vertical red line marks the true value of $f_{\text{NL}}^{\text{loc}}$ parameter used in simulations.}
\label{fig:comp_fnl_wout_photoz_errros}
\end{figure}

\section{Full likelihood plots}

\begin{figure}[ht!]
\plotone{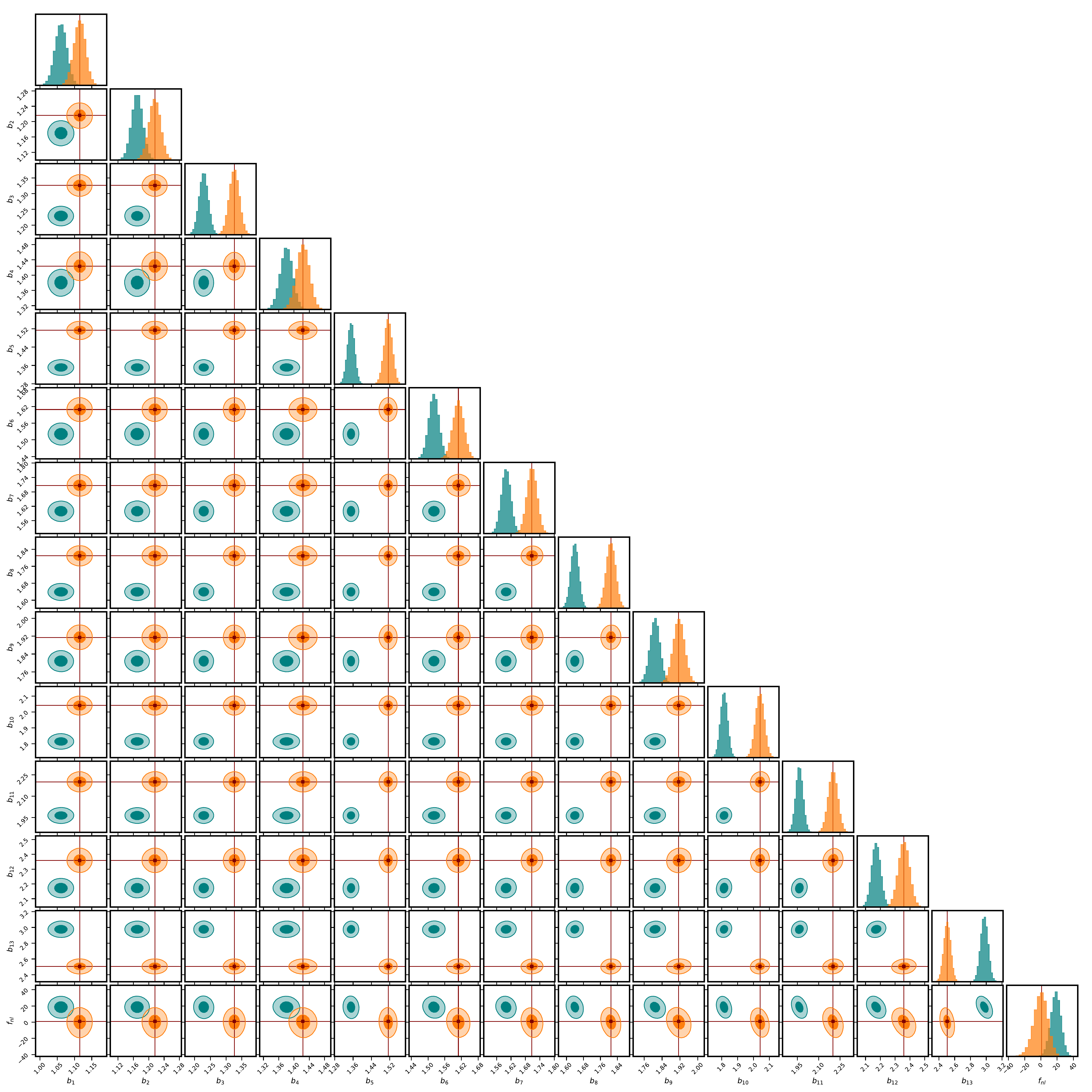}
\caption{Parameter posteriors obtained from maximum likelihood estimation for $f_{\text{NL}}^{\text{loc,true}} = 1$. The green histograms are the posteriors obtained following the convolution method to account for photo-$z$ errors, while the orange histograms are from the scattering matrix approach. The red lines are the true values of parameters used in simulations.}
\label{fig:full_likelihood_fnl_1}
\end{figure}

\begin{figure}[ht!]
\plotone{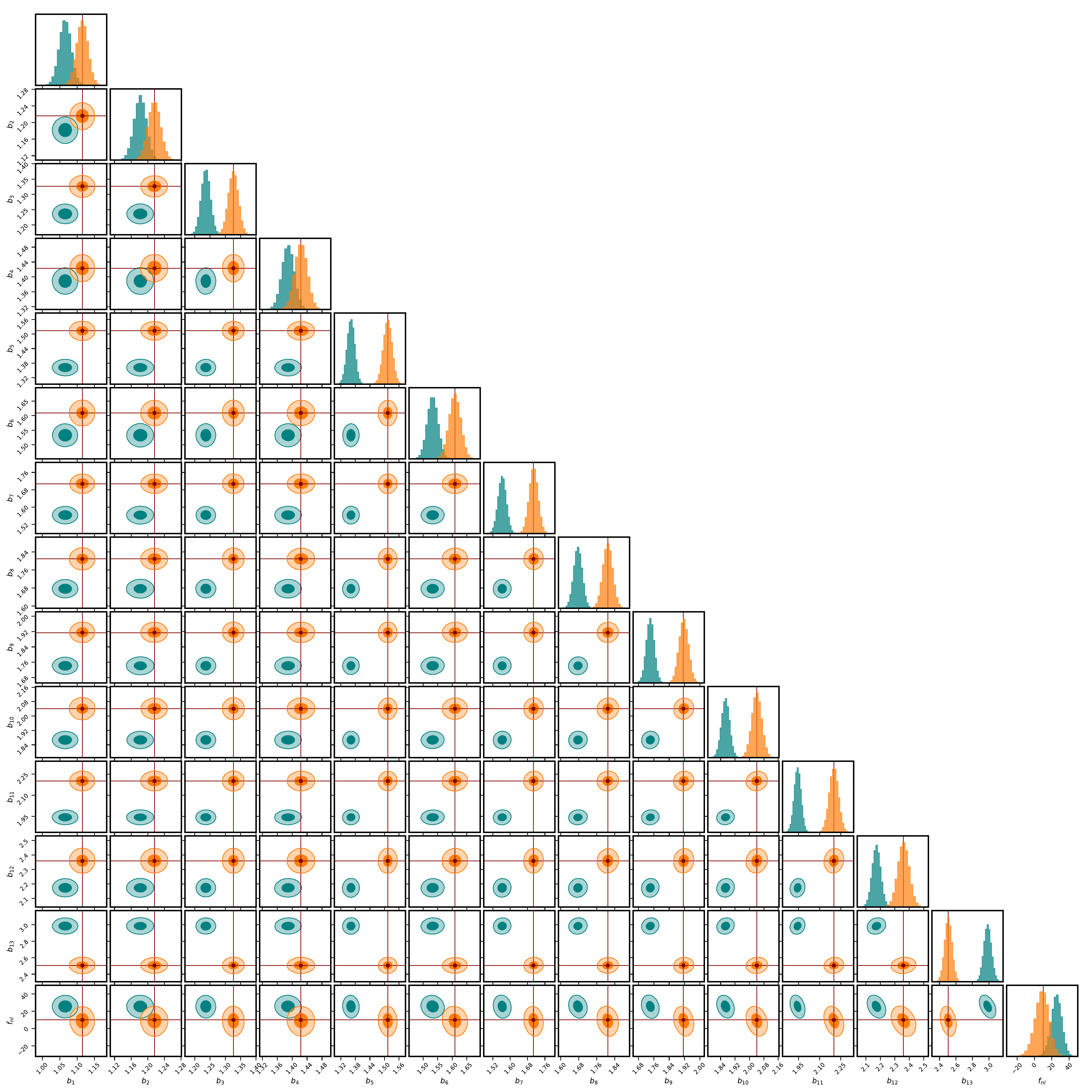}
\caption{Same as Figure \ref{fig:full_likelihood_fnl_1} but for $f_{\text{NL}}^{\text{loc,true}} = 10$}
\label{fig:full_likelihood_fnl_10}
\end{figure}

\begin{figure}[ht!]
\plotone{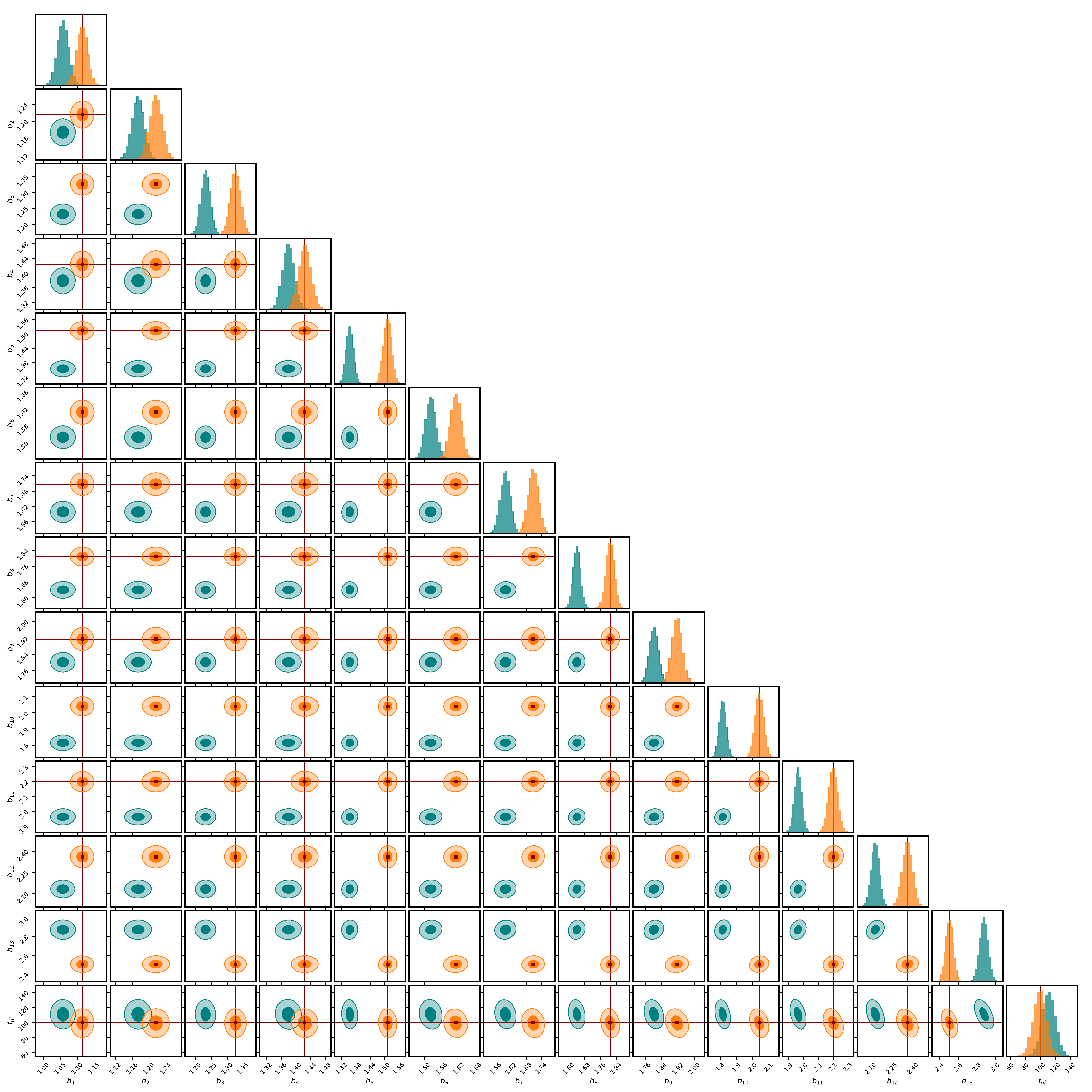}
\caption{Same as Figure \ref{fig:full_likelihood_fnl_1} but for $f_{\text{NL}}^{\text{loc,true}} = 100$}
\label{fig:full_likelihood_fnl_100}
\end{figure}

\bibliography{fnl_with_spherex}{}

\begin{thebibliography}{}
\expandafter\ifx\csname natexlab\endcsname\relax\def\natexlab#1{#1}\fi
\providecommand{\url}[1]{\href{#1}{#1}}
\providecommand{\dodoi}[1]{doi:~\href{http://doi.org/#1}{\nolinkurl{#1}}}
\providecommand{\doeprint}[1]{\href{http://ascl.net/#1}{\nolinkurl{http://ascl.net/#1}}}
\providecommand{\doarXiv}[1]{\href{https://arxiv.org/abs/#1}{\nolinkurl{https://arxiv.org/abs/#1}}}

\bibitem[{{Abazajian} {et~al.}(2016){Abazajian}, {Adshead}, {Ahmed}, {Allen}, {Alonso}, {Arnold}, {Baccigalupi}, {Bartlett}, {Battaglia}, {Benson}, {Bischoff}, {Borrill}, {Buza}, {Calabrese}, {Caldwell}, {Carlstrom}, {Chang}, {Crawford}, {Cyr-Racine}, {De Bernardis}, {de Haan}, {di Serego Alighieri}, {Dunkley}, {Dvorkin}, {Errard}, {Fabbian}, {Feeney}, {Ferraro}, {Filippini}, {Flauger}, {Fuller}, {Gluscevic}, {Green}, {Grin}, {Grohs}, {Henning}, {Hill}, {Hlozek}, {Holder}, {Holzapfel}, {Hu}, {Huffenberger}, {Keskitalo}, {Knox}, {Kosowsky}, {Kovac}, {Kovetz}, {Kuo}, {Kusaka}, {Le Jeune}, {Lee}, {Lilley}, {Loverde}, {Madhavacheril}, {Mantz}, {Marsh}, {McMahon}, {Meerburg}, {Meyers}, {Miller}, {Munoz}, {Nguyen}, {Niemack}, {Peloso}, {Peloton}, {Pogosian}, {Pryke}, {Raveri}, {Reichardt}, {Rocha}, {Rotti}, {Schaan}, {Schmittfull}, {Scott}, {Sehgal}, {Shandera}, {Sherwin}, {Smith}, {Sorbo}, {Starkman}, {Story}, {van Engelen}, {Vieira}, {Watson}, {Whitehorn}, \& {Kimmy Wu}}]{2016arXiv161002743A}
{Abazajian}, K.~N., {Adshead}, P., {Ahmed}, Z., {et~al.} 2016, arXiv e-prints, arXiv:1610.02743, \dodoi{10.48550/arXiv.1610.02743}

\bibitem[{{Alonso} {et~al.}(2019){Alonso}, {Sanchez}, {Slosar}, \& {LSST Dark Energy Science Collaboration}}]{2019MNRAS.484.4127A}
{Alonso}, D., {Sanchez}, J., {Slosar}, A., \& {LSST Dark Energy Science Collaboration}. 2019, \mnras, 484, 4127, \dodoi{10.1093/mnras/stz093}

\bibitem[{{Bartolo} {et~al.}(2004){Bartolo}, {Komatsu}, {Matarrese}, \& {Riotto}}]{2004PhR...402..103B}
{Bartolo}, N., {Komatsu}, E., {Matarrese}, S., \& {Riotto}, A. 2004, \physrep, 402, 103, \dodoi{10.1016/j.physrep.2004.08.022}

\bibitem[{{Cabass} {et~al.}(2022){Cabass}, {Ivanov}, {Philcox}, {Simonovi{\'c}}, \& {Zaldarriaga}}]{2022PhRvD.106d3506C}
{Cabass}, G., {Ivanov}, M.~M., {Philcox}, O. H.~E., {Simonovi{\'c}}, M., \& {Zaldarriaga}, M. 2022, \prd, 106, 043506, \dodoi{10.1103/PhysRevD.106.043506}

\bibitem[{{Creminelli} \& {Zaldarriaga}(2004)}]{2004JCAP...10..006C}
{Creminelli}, P., \& {Zaldarriaga}, M. 2004, \jcap, 2004, 006, \dodoi{10.1088/1475-7516/2004/10/006}

\bibitem[{{Dalal} {et~al.}(2008){Dalal}, {Dor{\'e}}, {Huterer}, \& {Shirokov}}]{2008PhRvD..77l3514D}
{Dalal}, N., {Dor{\'e}}, O., {Huterer}, D., \& {Shirokov}, A. 2008, \prd, 77, 123514, \dodoi{10.1103/PhysRevD.77.123514}

\bibitem[{{D'Amico} {et~al.}(2022){D'Amico}, {Lewandowski}, {Senatore}, \& {Zhang}}]{2022arXiv220111518D}
{D'Amico}, G., {Lewandowski}, M., {Senatore}, L., \& {Zhang}, P. 2022, arXiv e-prints, arXiv:2201.11518, \dodoi{10.48550/arXiv.2201.11518}

\bibitem[{{DESI Collaboration} {et~al.}(2016){DESI Collaboration}, {Aghamousa}, {Aguilar}, {Ahlen}, {Alam}, {Allen}, {Allende Prieto}, {Annis}, {Bailey}, {Balland}, {Ballester}, {Baltay}, {Beaufore}, {Bebek}, {Beers}, {Bell}, {Bernal}, {Besuner}, {Beutler}, {Blake}, {Bleuler}, {Blomqvist}, {Blum}, {Bolton}, {Briceno}, {Brooks}, {Brownstein}, {Buckley-Geer}, {Burden}, {Burtin}, {Busca}, {Cahn}, {Cai}, {Cardiel-Sas}, {Carlberg}, {Carton}, {Casas}, {Castander}, {Cervantes-Cota}, {Claybaugh}, {Close}, {Coker}, {Cole}, {Comparat}, {Cooper}, {Cousinou}, {Crocce}, {Cuby}, {Cunningham}, {Davis}, {Dawson}, {de la Macorra}, {De Vicente}, {Delubac}, {Derwent}, {Dey}, {Dhungana}, {Ding}, {Doel}, {Duan}, {Ealet}, {Edelstein}, {Eftekharzadeh}, {Eisenstein}, {Elliott}, {Escoffier}, {Evatt}, {Fagrelius}, {Fan}, {Fanning}, {Farahi}, {Farihi}, {Favole}, {Feng}, {Fernandez}, {Findlay}, {Finkbeiner}, {Fitzpatrick}, {Flaugher}, {Flender}, {Font-Ribera}, {Forero-Romero}, {Fosalba}, {Frenk}, {Fumagalli}, {Gaensicke}, {Gallo},
  {Garcia-Bellido}, {Gaztanaga}, {Pietro Gentile Fusillo}, {Gerard}, {Gershkovich}, {Giannantonio}, {Gillet}, {Gonzalez-de-Rivera}, {Gonzalez-Perez}, {Gott}, {Graur}, {Gutierrez}, {Guy}, {Habib}, {Heetderks}, {Heetderks}, {Heitmann}, {Hellwing}, {Herrera}, {Ho}, {Holland}, {Honscheid}, {Huff}, {Hutchinson}, {Huterer}, {Hwang}, {Illa Laguna}, {Ishikawa}, {Jacobs}, {Jeffrey}, {Jelinsky}, {Jennings}, {Jiang}, {Jimenez}, {Johnson}, {Joyce}, {Jullo}, {Juneau}, {Kama}, {Karcher}, {Karkar}, {Kehoe}, {Kennamer}, {Kent}, {Kilbinger}, {Kim}, {Kirkby}, {Kisner}, {Kitanidis}, {Kneib}, {Koposov}, {Kovacs}, {Koyama}, {Kremin}, {Kron}, {Kronig}, {Kueter-Young}, {Lacey}, {Lafever}, {Lahav}, {Lambert}, {Lampton}, {Landriau}, {Lang}, {Lauer}, {Le Goff}, {Le Guillou}, {Le Van Suu}, {Lee}, {Lee}, {Leitner}, {Lesser}, {Levi}, {L'Huillier}, {Li}, {Liang}, {Lin}, {Linder}, {Loebman}, {Luki{\'c}}, {Ma}, {MacCrann}, {Magneville}, {Makarem}, {Manera}, {Manser}, {Marshall}, {Martini}, {Massey}, {Matheson}, {McCauley}, {McDonald},
  {McGreer}, {Meisner}, {Metcalfe}, {Miller}, {Miquel}, {Moustakas}, {Myers}, {Naik}, {Newman}, {Nichol}, {Nicola}, {Nicolati da Costa}, {Nie}, {Niz}, {Norberg}, {Nord}, {Norman}, {Nugent}, {O'Brien}, {Oh}, {Olsen}, {Padilla}, {Padmanabhan}, {Padmanabhan}, {Palanque-Delabrouille}, {Palmese}, {Pappalardo}, {P{\^a}ris}, {Park}, {Patej}, {Peacock}, {Peiris}, {Peng}, {Percival}, {Perruchot}, {Pieri}, {Pogge}, {Pollack}, {Poppett}, {Prada}, {Prakash}, {Probst}, {Rabinowitz}, {Raichoor}, {Ree}, {Refregier}, {Regal}, {Reid}, {Reil}, {Rezaie}, {Rockosi}, {Roe}, {Ronayette}, {Roodman}, {Ross}, {Ross}, {Rossi}, {Rozo}, {Ruhlmann-Kleider}, {Rykoff}, {Sabiu}, {Samushia}, {Sanchez}, {Sanchez}, {Schlegel}, {Schneider}, {Schubnell}, {Secroun}, {Seljak}, {Seo}, {Serrano}, {Shafieloo}, {Shan}, {Sharples}, {Sholl}, {Shourt}, {Silber}, {Silva}, {Sirk}, {Slosar}, {Smith}, {Smoot}, {Som}, {Song}, {Sprayberry}, {Staten}, {Stefanik}, {Tarle}, {Sien Tie}, {Tinker}, {Tojeiro}, {Valdes}, {Valenzuela}, {Valluri}, {Vargas-Magana},
  {Verde}, {Walker}, {Wang}, {Wang}, {Weaver}, {Weaverdyck}, {Wechsler}, {Weinberg}, {White}, {Yang}, {Yeche}, {Zhang}, {Zhao}, {Zheng}, {Zhou}, {Zhou}, {Zhu}, {Zou}, \& {Zu}}]{2016arXiv161100036D}
{DESI Collaboration}, {Aghamousa}, A., {Aguilar}, J., {et~al.} 2016, arXiv e-prints, arXiv:1611.00036, \dodoi{10.48550/arXiv.1611.00036}

\bibitem[{{Dor{\'e}} {et~al.}(2014){Dor{\'e}}, {Bock}, {Ashby}, {Capak}, {Cooray}, {de Putter}, {Eifler}, {Flagey}, {Gong}, {Habib}, {Heitmann}, {Hirata}, {Jeong}, {Katti}, {Korngut}, {Krause}, {Lee}, {Masters}, {Mauskopf}, {Melnick}, {Mennesson}, {Nguyen}, {{\"O}berg}, {Pullen}, {Raccanelli}, {Smith}, {Song}, {Tolls}, {Unwin}, {Venumadhav}, {Viero}, {Werner}, \& {Zemcov}}]{2014arXiv1412.4872D}
{Dor{\'e}}, O., {Bock}, J., {Ashby}, M., {et~al.} 2014, arXiv e-prints, arXiv:1412.4872, \dodoi{10.48550/arXiv.1412.4872}

\bibitem[{{Euclid Collaboration} {et~al.}(2024){Euclid Collaboration}, {Mellier}, {Abdurro'uf}, {Acevedo Barroso}, {Ach{\'u}carro}, {Adamek}, {Adam}, {Addison}, {Aghanim}, {Aguena}, {Ajani}, {Akrami}, {Al-Bahlawan}, {Alavi}, {Albuquerque}, {Alestas}, {Alguero}, {Allaoui}, {Allen}, {Allevato}, {Alonso-Tetilla}, {Altieri}, {Alvarez-Candal}, {Alvi}, {Amara}, {Amendola}, {Amiaux}, {Andika}, {Andreon}, {Andrews}, {Angora}, {Angulo}, {Annibali}, {Anselmi}, {Anselmi}, {Arcari}, {Archidiacono}, {Aric{\`o}}, {Arnaud}, {Arnouts}, {Asgari}, {Asorey}, {Atayde}, {Atek}, {Atrio-Barandela}, {Aubert}, {Aubourg}, {Auphan}, {Auricchio}, {Aussel}, {Aussel}, {Avelino}, {Avgoustidis}, {Avila}, {Awan}, {Azzollini}, {Baccigalupi}, {Bachelet}, {Bacon}, {Baes}, {Bagley}, {Bahr-Kalus}, {Balaguera-Antolinez}, {Balbinot}, {Balcells}, {Baldi}, {Baldry}, {Balestra}, {Ballardini}, {Ballester}, {Balogh}, {Ba{\~n}ados}, {Barbier}, {Bardelli}, {Baron}, {Barreiro}, {Barrena}, {Barriere}, {Barros}, {Barthelemy}, {Bartolo}, {Basset},
  {Battaglia}, {Battisti}, {Baugh}, {Baumont}, {Bazzanini}, {Beaulieu}, {Beckmann}, {Belikov}, {Bel}, {Bellagamba}, {Bella}, {Bellini}, {Benabed}, {Bender}, {Benevento}, {Bennett}, {Benson}, {Bergamini}, {Bermejo-Climent}, {Bernardeau}, {Bertacca}, {Berthe}, {Berthier}, {Bethermin}, {Beutler}, {Bevillon}, {Bhargava}, {Bhatawdekar}, {Bianchi}, {Bisigello}, {Biviano}, {Blake}, {Blanchard}, {Blazek}, {Blot}, {Bosco}, {Bodendorf}, {Boenke}, {B{\"o}hringer}, {Boldrini}, {Bolzonella}, {Bonchi}, {Bonici}, {Bonino}, {Bonino}, {Bonvin}, {Bon}, {Booth}, {Borgani}, {Borlaff}, {Borsato}, {Bosco}, {Bose}, {Botticella}, {Boucaud}, {Bouche}, {Boucher}, {Boutigny}, {Bouvard}, {Bouwens}, {Bouy}, {Bowler}, {Bozza}, {Bozzo}, {Branchini}, {Brando}, {Brau-Nogue}, {Brekke}, {Bremer}, {Brescia}, {Breton}, {Brinchmann}, {Brinckmann}, {Brockley-Blatt}, {Brodwin}, {Brouard}, {Brown}, {Bruton}, {Bucko}, {Buddelmeijer}, {Buenadicha}, {Buitrago}, {Burger}, {Burigana}, {Busillo}, {Busonero}, {Cabanac}, {Cabayol-Garcia}, {Cagliari},
  {Caillat}, {Caillat}, {Calabrese}, {Calabro}, {Calderone}, {Calura}, {Camacho Quevedo}, {Camera}, {Campos}, {Canas-Herrera}, {Candini}, {Cantiello}, {Capobianco}, {Cappellaro}, {Cappelluti}, {Cappi}, {Caputi}, {Cara}, {Carbone}, {Cardone}, {Carella}, {Carlberg}, {Carle}, {Carminati}, {Caro}, {Carrasco}, {Carretero}, {Carrilho}, {Carron Duque}, {Carry}, {Carvalho}, {Carvalho}, {Casas}, {Casas}, {Casenove}, {Casey}, {Cassata}, {Castander}, {Castelao}, {Castellano}, {Castiblanco}, {Castignani}, {Castro}, {Cavet}, {Cavuoti}, {Chabaud}, {Chambers}, {Charles}, {Charlot}, {Chartab}, {Chary}, {Chaumeil}, {Cho}, {Chon}, {Ciancetta}, {Ciliegi}, {Cimatti}, {Cimino}, {Cioni}, {Claydon}, {Cleland}, {Cl{\'e}ment}, {Clements}, {Clerc}, {Clesse}, {Codis}, {Cogato}, {Colbert}, {Cole}, {Coles}, {Collett}, {Collins}, {Colodro-Conde}, {Colombo}, {Combes}, {Conforti}, {Congedo}, {Conseil}, {Conselice}, {Contarini}, {Contini}, {Conversi}, {Cooray}, {Copin}, {Corasaniti}, {Corcho-Caballero}, {Corcione}, {Cordes}, {Corpace},
  {Correnti}, {Costanzi}, {Costille}, {Courbin}, {Courcoult Mifsud}, {Courtois}, {Cousinou}, {Covone}, {Cowell}, {Cragg}, {Cresci}, {Cristiani}, {Crocce}, {Cropper}, {E Crouzet}, {Csizi}, {Cuby}, {Cucchetti}, {Cucciati}, {Cuillandre}, {Cunha}, {Cuozzo}, {Daddi}, {D'Addona}, {Dafonte}, {Dagoneau}, {Dalessandro}, {Dalton}, {D'Amico}, {Dannerbauer}, {Danto}, {Das}, {Da Silva}, {da Silva}, {d'Assignies Doumerg}, {Daste}, {Davies}, {Davini}, {Dayal}, {de Boer}, {Decarli}, {De Caro}, {Degaudenzi}, {Degni}, {de Jong}, {de la Bella}, {de la Torre}, {Delhaise}, {Delley}, {Delucchi}, {De Lucia}, {Denniston}, {De Paolis}, {De Petris}, {Derosa}, {Desai}, {Desjacques}, {Despali}, {Desprez}, {De Vicente-Albendea}, {Deville}, {Dias}, {D{\'\i}az-S{\'a}nchez}, {Diaz}, {Di Domizio}, {Diego}, {Di Ferdinando}, {Di Giorgio}, {Dimauro}, {Dinis}, {Dolag}, {Dolding}, {Dole}, {Dom{\'\i}nguez S{\'a}nchez}, {Dor{\'e}}, {Dournac}, {Douspis}, {Dreihahn}, {Droge}, {Dryer}, {Dubath}, {Duc}, {Ducret}, {Duffy}, {Dufresne}, {Duncan}, {Dupac},
  {Duret}, {Durrer}, {Durret}, {Dusini}, {Ealet}, {Eggemeier}, {Eisenhardt}, {Elbaz}, {Elkhashab}, {Ellien}, {Endicott}, {Enia}, {Erben}, {Escartin Vigo}, {Escoffier}, {Escudero Sanz}, {Essert}, {Ettori}, {Ezziati}, {Fabbian}, {Fabricius}, {Fang}, {Farina}, {Farina}, {Farinelli}, {Farrens}, {Faustini}, {Feltre}, {Ferguson}, {Ferrando}, {Ferrari}, {Ferr{\'e}-Mateu}, {Ferreira}, {Ferreras}, {Ferrero}, {Ferriol}, {Ferruit}, {Filleul}, {Finelli}, {Finkelstein}, {Finoguenov}, {Fiorini}, {Flentge}, {Focardi}, {Fonseca}, {Fontana}, {Fontanot}, {Fornari}, {Fosalba}, {Fossati}, {Fotopoulou}, {Fouchez}, {Fourmanoit}, {Frailis}, {Fraix-Burnet}, {Franceschi}, {Franco}, {Franzetti}, {Freihoefer}, {Frenk}, {Frittoli}, {Frugier}, {Frusciante}, {Fumagalli}, {Fumagalli}, {Fumana}, {Fu}, {Gabarra}, {Galeotta}, {Galluccio}, {Ganga}, {Gao}, {Garc{\'\i}a-Bellido}, {Garcia}, {Gardner}, {Garilli}, {Gaspar-Venancio}, {Gasparetto}, {Gautard}, {Gavazzi}, {Gaztanaga}, {Genolet}, {Genova Santos}, {Gentile}, {George}, {Gerbino},
  {Ghaffari}, {Giacomini}, {Gianotti}, {Gibb}, {Gillard}, {Gillis}, {Ginolfi}, {Giocoli}, {Girardi}, {Giri}, {Goh}, {G{\'o}mez-Alvarez}, {Gonzalez-Perez}, {Gonzalez}, {Gonzalez}, {Gonzalez}, {Gouyou Beauchamps}, {Gozaliasl}, {Gracia-Carpio}, {Grandis}, {Granett}, {Granvik}, {Grazian}, {Gregorio}, {Grenet}, {Grillo}, {Grupp}, {Gruppioni}, {Gruppuso}, {Guerbuez}, {Guerrini}, {Guidi}, {Guillard}, {Gutierrez}, {Guttridge}, {Guzzo}, {Gwyn}, {Haapala}, {Haase}, {Haddow}, {Hailey}, {Hall}, {Hall}, {Hamaus}, {Haridasu}, {Harnois-D{\'e}raps}, {Harper}, {Hartley}, {Hasinger}, {Hassani}, {Hatch}, {Haugan}, {H{\"a}u{\ss}ler}, {Heavens}, {Heisenberg}, {Helmi}, {Helou}, {Hemmati}, {Henares}, {Herent}, {Hern{\'a}ndez-Monteagudo}, {Heuberger}, {Hewett}, {Heydenreich}, {Hildebrandt}, {Hirschmann}, {Hjorth}, {Hoar}, {Hoekstra}, {Holland}, {Holliman}, {Holmes}, {Hook}, {Horeau}, {Hormuth}, {Hornstrup}, {Hosseini}, {Hu}, {Hudelot}, {Hudson}, {Huertas-Company}, {Huff}, {Hughes}, {Humphrey}, {Hunt}, {Huynh}, {Ibata}, {Ichikawa},
  {Iglesias-Groth}, {Ilbert}, {Ili{\'c}}, {Ingoglia}, {Iodice}, {Israel}, {Israelsson}, {Izzo}, {Jablonka}, {Jackson}, {Jacobson}, {Jafariyazani}, {Jahnke}, {Jain}, {Jansen}, {Jarvis}, {Jasche}, {Jauzac}, {Jeffrey}, {Jhabvala}, {Jimenez-Teja}, {Jimenez Mu{\~n}oz}, {Joachimi}, {Johansson}, {Joudaki}, {Jullo}, {Kajava}, {Kang}, {Kannawadi}, {Kansal}, {Karagiannis}, {K{\"a}rcher}, {Kashlinsky}, {Kazandjian}, {Keck}, {Keih{\"a}nen}, {Kerins}, {Kermiche}, {Khalil}, {Kiessling}, {Kiiveri}, {Kilbinger}, {Kim}, {King}, {Kirkpatrick}, {Kitching}, {Kluge}, {Knabenhans}, {Knapen}, {Knebe}, {Kneib}, {Kohley}, {Koopmans}, {Koskinen}, {Koulouridis}, {Kou}, {Kov{\'a}cs}, {Kova{\v{c}}i{\'c}}, {Kowalczyk}, {Koyama}, {Kraljic}, {Krause}, {Kruk}, {Kubik}, {Kuchner}, {Kuijken}, {K{\"u}mmel}, {Kunz}, {Kurki-Suonio}, {Lacasa}, {Lacey}, {La Franca}, {Lagarde}, {Lahav}, {Laigle}, {La Marca}, {La Marle}, {Lamine}, {Lam}, {Lan{\c{c}}on}, {Landt}, {Langer}, {Lapi}, {Larcheveque}, {Larsen}, {Lattanzi}, {Laudisio}, {Laugier}, {Laureijs},
  {Laurent}, {Lavaux}, {Lawrenson}, {Lazanu}, {Lazeyras}, {Le Boulc'h}, {Le Brun}, {Le Brun}, {Leclercq}, {Lee}, {Le Graet}, {Legrand}, {Leirvik}, {Le Jeune}, {Lembo}, {Le Mignant}, {Lepinzan}, {Lepori}, {Le Reun}, {Leroy}, {Lesci}, {Lesgourgues}, {Leuzzi}, {Levi}, {Liaudat}, {Libet}, {Liebing}, {Ligori}, {Lilje}, {Lin}, {Linde}, {Linder}, {Lindholm}, {Linke}, {Li}, {Liu}, {Lloro}, {Lobo}, {Lodieu}, {Lombardi}, {Lombriser}, {Lonare}, {Longo}, {L{\'o}pez-Caniego}, {Lopez Lopez}, {Alvarez}, {Loureiro}, {Loveday}, {Lusso}, {Macias-Perez}, {Maciaszek}, {Maggio}, {Magliocchetti}, {Magnard}, {Magnier}, {Magro}, {Mahler}, {Mainetti}, {Maino}, {Maiorano}, {Maiorano}, {Malavasi}, {Mamon}, {Mancini}, {Mandelbaum}, {Manera}, {Manj{\'o}n-Garc{\'\i}a}, {Mannucci}, {Mansutti}, {Manteiga Outeiro}, {Maoli}, {Maraston}, {Marcin}, {Marcos-Arenal}, {Margalef-Bentabol}, {Marggraf}, {Marinucci}, {Marinucci}, {Markovic}, {Marleau}, {Marpaud}, {Martignac}, {Mart{\'\i}n-Fleitas}, {Martin-Moruno}, {Martin}, {Martinelli}, {Martinet},
  {Martin}, {Martins}, {Marulli}, {Massari}, {Massey}, {Masters}, {Matarrese}, {Matsuoka}, {Matthew}, {Maughan}, {Mauri}, {Maurin}, {Maurogordato}, {McCarthy}, {McConnachie}, {McCracken}, {McDonald}, {McEwen}, {McPartland}, {Medinaceli}, {Mehta}, {Mei}, {Melchior}, {Melin}, {M{\'e}nard}, {Mendes}, {Mendez-Abreu}, {Meneghetti}, {Mercurio}, {Merlin}, {Metcalf}, {Meylan}, {Migliaccio}, {Mignoli}, {Miller}, {Miluzio}, {Milvang-Jensen}, {Mimoso}, {Miquel}, {Miyatake}, {Mobasher}, {Mohr}, {Monaco}, {Mongui{\'o}}, {Montoro}, {Mora}, {Moradinezhad Dizgah}, {Moresco}, {Moretti}, {Morgante}, {Morisset}, {Moriya}, {Morris}, {Mortlock}, {Moscardini}, {Mota}, {Mottet}, {Moustakas}, {Moutard}, {M{\"u}ller}, {Munari}, {Murphree}, {Murray}, {Murray}, {Musi}, {Nadathur}, {Nagam}, {Nagao}, {Naidoo}, {Nakajima}, {Nally}, {Natoli}, {Navarro-Alsina}, {Navarro Girones}, {Neissner}, {Nersesian}, {Nesseris}, {Nguyen-Kim}, {Nicastro}, {Nichol}, {Nielbock}, {Niemi}, {Nieto}, {Nilsson}, {Noller}, {Norberg}, {Nouri-Zonoz}, {Ntelis},
  {Nucita}, {Nugent}, {Nunes}, {Nutma}, {Ocampo}, {Odier}, {Oesch}, {Oguri}, {Magalhaes Oliveira}, {Onoue}, {Oosterbroek}, {Oppizzi}, {Ordenovic}, {Osato}, {Pacaud}, {Pace}, {Padilla}, {Paech}, {Pagano}, {Page}, {Palazzi}, {Paltani}, {Pamuk}, {Pandolfi}, {Paoletti}, {Paolillo}, {Papaderos}, {Pardede}, {Parimbelli}, {Parmar}, {Partmann}, {Pasian}, {Passalacqua}, {Paterson}, {Patrizii}, {Pattison}, {Paulino-Afonso}, {Paviot}, {Peacock}, {Pearce}, {Pedersen}, {Peel}, {Peletier}, {Pellejero Ibanez}, {Pello}, {Penny}, {Percival}, {Perez-Garrido}, {Perotto}, {Pettorino}, {Pezzotta}, {Pezzuto}, {Philippon}, {Pierre}, {Piersanti}, {Pietroni}, {Piga}, {Pilo}, {Pires}, {Pisani}, {Pizzella}, {Pizzuti}, {Plana}, {Polenta}, {Pollack}, {Poncet}, {P{\"o}ntinen}, {Pool}, {Popa}, {Popa}, {Popp}, {Porciani}, {Porth}, {Potter}, {Poulain}, {Pourtsidou}, {Pozzetti}, {Prandoni}, {Pratt}, {Prezelus}, {Prieto}, {Pugno}, {Quai}, {Quilley}, {Racca}, {Raccanelli}, {R{\'a}cz}, {Radinovi{\'c}}, {Radovich}, {Ragagnin}, {Ragnit}, {Raison},
  {Ramos-Chernenko}, {Ranc}, {Rasera}, {Raylet}, {Rebolo}, {Refregier}, {Reimberg}, {Reiprich}, {Renk}, {Renzi}, {Retre}, {Revaz}, {Reyl{\'e}}, {Reynolds}, {Rhodes}, {Ricci}, {Ricci}, {Riccio}, {Ricken}, {Rissanen}, {Risso}, {Rix}, {Robin}, {Rocca-Volmerange}, {Rocci}, {Rodenhuis}, {Rodighiero}, {Rodriguez Monroy}, {Rollins}, {Romanello}, {Roman}, {Romelli}, {Romero-Gomez}, {Roncarelli}, {Rosati}, {Rosset}, {Rossetti}, {Roster}, {Rottgering}, {Rozas-Fern{\'a}ndez}, {Ruane}, {Rubino-Martin}, {Rudolph}, {Ruppin}, {Rusholme}, {Sacquegna}, {S{\'a}ez-Casares}, {Saga}, {Saglia}, {Sahl{\'e}n}, {Saifollahi}, {Sakr}, {Salvalaggio}, {Salvaterra}, {Salvati}, {Salvato}, {Salvignol}, {S{\'a}nchez}, {Sanchez}, {Sanders}, {Sapone}, {Saponara}, {Sarpa}, {Sarron}, {Sartori}, {Sartoris}, {Sassolas}, {Sauniere}, {Sauvage}, {Sawicki}, {Scaramella}, {Scarlata}, {Scharr{\'e}}, {Schaye}, {Schewtschenko}, {Schindler}, {Schinnerer}, {Schirmer}, {Schmidt}, {Schmidt}, {Schmidt}, {Schneider}, {Schneider}, {Schneider}, {Sch{\"o}neberg},
  {Schrabback}, {Schultheis}, {Schulz}, {Schuster}, {Schwartz}, {Sciotti}, {Scodeggio}, {Scognamiglio}, {Scott}, {Scottez}, {Secroun}, {Sefusatti}, {Seidel}, {Seiffert}, {Sellentin}, {Selwood}, {Semboloni}, {Sereno}, {Serjeant}, {Serrano}, {Setnikar}, {Shankar}, {Sharples}, {Short}, {Shulevski}, {Shuntov}, {Sias}, {Sikkema}, {Silvestri}, {Simon}, {Sirignano}, {Sirri}, {Skottfelt}, {Slezak}, {Sluse}, {Smith}, {Smith}, {Smith}, {Smit}, {Soldano}, {Solheim}, {Sorce}, {Sorrenti}, {Soubrie}, {Spinoglio}, {Spurio Mancini}, {Stadel}, {Stagnaro}, {Stanco}, {Stanford}, {Starck}, {Stassi}, {Steinwagner}, {Stern}, {Stone}, {Strada}, {Strafella}, {Stramaccioni}, {Surace}, {Sureau}, {Suyu}, {Swindells}, {Szafraniec}, {Szapudi}, {Taamoli}, {Talia}, {Tallada-Cresp{\'\i}}, {Tanidis}, {Tao}, {Tarr{\'\i}o}, {Tavagnacco}, {Taylor}, {Taylor}, {Taylor}, {Teixeira}, {Tenti}, {Teodoro Idiago}, {Teplitz}, {Tereno}, {Tessore}, {Testa}, {Testera}, {Tewes}, {Teyssier}, {Theret}, {Thizy}, {Thomas}, {Toba}, {Toft}, {Toledo-Moreo},
  {Tolstoy}, {Tommasi}, {Torbaniuk}, {Torradeflot}, {Tortora}, {Tosi}, {Tosti}, {Trifoglio}, {Troja}, {Trombetti}, {Tronconi}, {Tsedrik}, {Tsyganov}, {Tucci}, {Tutusaus}, {Uhlemann}, {Ulivi}, {Urbano}, {Vacher}, {Vaillon}, {Valageas}, {Valdes}, {Valentijn}, {Valenziano}, {Valieri}, {Valiviita}, {Van den Broeck}, {Vassallo}, {Vavrek}, {Vega-Ferrero}, {Venemans}, {Venhola}, {Ventura}, {Verdoes Kleijn}, {Vergani}, {Verma}, {Vernizzi}, {Veropalumbo}, {Verza}, {Vescovi}, {Vibert}, {Viel}, {Vielzeuf}, {Viglione}, {Viitanen}, {Villaescusa-Navarro}, {Vinciguerra}, {Visticot}, {Voggel}, {von Wietersheim-Kramsta}, {Vriend}, {Wachter}, {Walmsley}, {Walth}, {Walton}, {Walton}, {Wander}, {Wang}, {Wang}, {Weaver}, {Weller}, {Wetzstein}, {Whalen}, {Whittam}, {Widmer}, {Wiesmann}, {Wilde}, {Williams}, {Winther}, {Wittje}, {Wong}, {Wright}, {Yankelevich}, {Yeung}, {Yoon}, {Youles}, {Yung}, {Zacchei}, {Zalesky}, {Zamorani}, {Zamorano Vitorelli}, {Zanoni Marc}, {Zennaro}, {Zerbi}, {Zinchenko}, {Zoubian}, {Zucca}, \&
  {Zumalacarregui}}]{2024arXiv240513491E}
{Euclid Collaboration}, {Mellier}, Y., {Abdurro'uf}, {et~al.} 2024, arXiv e-prints, arXiv:2405.13491, \dodoi{10.48550/arXiv.2405.13491}

\bibitem[{{Fonseca} {et~al.}(2017){Fonseca}, {Maartens}, \& {Santos}}]{2017MNRAS.466.2780F}
{Fonseca}, J., {Maartens}, R., \& {Santos}, M.~G. 2017, \mnras, 466, 2780, \dodoi{10.1093/mnras/stw3248}

\bibitem[{{Fonseca} {et~al.}(2018){Fonseca}, {Maartens}, \& {Santos}}]{2018MNRAS.479.3490F}
---. 2018, \mnras, 479, 3490, \dodoi{10.1093/mnras/sty1702}

\bibitem[{{Foreman-Mackey} {et~al.}(2013){Foreman-Mackey}, {Hogg}, {Lang}, \& {Goodman}}]{2013PASP..125..306F}
{Foreman-Mackey}, D., {Hogg}, D.~W., {Lang}, D., \& {Goodman}, J. 2013, \pasp, 125, 306, \dodoi{10.1086/670067}

\bibitem[{{Giannantonio} {et~al.}(2014){Giannantonio}, {Ross}, {Percival}, {Crittenden}, {Bacher}, {Kilbinger}, {Nichol}, \& {Weller}}]{2014PhRvD..89b3511G}
{Giannantonio}, T., {Ross}, A.~J., {Percival}, W.~J., {et~al.} 2014, \prd, 89, 023511, \dodoi{10.1103/PhysRevD.89.023511}

\bibitem[{{Gomes} {et~al.}(2020){Gomes}, {Camera}, {Jarvis}, {Hale}, \& {Fonseca}}]{2020MNRAS.492.1513G}
{Gomes}, Z., {Camera}, S., {Jarvis}, M.~J., {Hale}, C., \& {Fonseca}, J. 2020, \mnras, 492, 1513, \dodoi{10.1093/mnras/stz3581}

\bibitem[{{G{\'o}rski} {et~al.}(2005){G{\'o}rski}, {Hivon}, {Banday}, {Wandelt}, {Hansen}, {Reinecke}, \& {Bartelmann}}]{2005ApJ...622..759G}
{G{\'o}rski}, K.~M., {Hivon}, E., {Banday}, A.~J., {et~al.} 2005, \apj, 622, 759, \dodoi{10.1086/427976}

\bibitem[{{Guth}(1981)}]{1981PhRvD..23..347G}
{Guth}, A.~H. 1981, \prd, 23, 347, \dodoi{10.1103/PhysRevD.23.347}

\bibitem[{{Hivon} {et~al.}(2002){Hivon}, {G{\'o}rski}, {Netterfield}, {Crill}, {Prunet}, \& {Hansen}}]{2002ApJ...567....2H}
{Hivon}, E., {G{\'o}rski}, K.~M., {Netterfield}, C.~B., {et~al.} 2002, \apj, 567, 2, \dodoi{10.1086/338126}

\bibitem[{{Ivezi{\'c}} {et~al.}(2019){Ivezi{\'c}}, {Kahn}, {Tyson}, {Abel}, {Acosta}, {Allsman}, {Alonso}, {AlSayyad}, {Anderson}, {Andrew}, {Angel}, {Angeli}, {Ansari}, {Antilogus}, {Araujo}, {Armstrong}, {Arndt}, {Astier}, {Aubourg}, {Auza}, {Axelrod}, {Bard}, {Barr}, {Barrau}, {Bartlett}, {Bauer}, {Bauman}, {Baumont}, {Bechtol}, {Bechtol}, {Becker}, {Becla}, {Beldica}, {Bellavia}, {Bianco}, {Biswas}, {Blanc}, {Blazek}, {Blandford}, {Bloom}, {Bogart}, {Bond}, {Booth}, {Borgland}, {Borne}, {Bosch}, {Boutigny}, {Brackett}, {Bradshaw}, {Brandt}, {Brown}, {Bullock}, {Burchat}, {Burke}, {Cagnoli}, {Calabrese}, {Callahan}, {Callen}, {Carlin}, {Carlson}, {Chandrasekharan}, {Charles-Emerson}, {Chesley}, {Cheu}, {Chiang}, {Chiang}, {Chirino}, {Chow}, {Ciardi}, {Claver}, {Cohen-Tanugi}, {Cockrum}, {Coles}, {Connolly}, {Cook}, {Cooray}, {Covey}, {Cribbs}, {Cui}, {Cutri}, {Daly}, {Daniel}, {Daruich}, {Daubard}, {Daues}, {Dawson}, {Delgado}, {Dellapenna}, {de Peyster}, {de Val-Borro}, {Digel}, {Doherty}, {Dubois},
  {Dubois-Felsmann}, {Durech}, {Economou}, {Eifler}, {Eracleous}, {Emmons}, {Fausti Neto}, {Ferguson}, {Figueroa}, {Fisher-Levine}, {Focke}, {Foss}, {Frank}, {Freemon}, {Gangler}, {Gawiser}, {Geary}, {Gee}, {Geha}, {Gessner}, {Gibson}, {Gilmore}, {Glanzman}, {Glick}, {Goldina}, {Goldstein}, {Goodenow}, {Graham}, {Gressler}, {Gris}, {Guy}, {Guyonnet}, {Haller}, {Harris}, {Hascall}, {Haupt}, {Hernandez}, {Herrmann}, {Hileman}, {Hoblitt}, {Hodgson}, {Hogan}, {Howard}, {Huang}, {Huffer}, {Ingraham}, {Innes}, {Jacoby}, {Jain}, {Jammes}, {Jee}, {Jenness}, {Jernigan}, {Jevremovi{\'c}}, {Johns}, {Johnson}, {Johnson}, {Jones}, {Juramy-Gilles}, {Juri{\'c}}, {Kalirai}, {Kallivayalil}, {Kalmbach}, {Kantor}, {Karst}, {Kasliwal}, {Kelly}, {Kessler}, {Kinnison}, {Kirkby}, {Knox}, {Kotov}, {Krabbendam}, {Krughoff}, {Kub{\'a}nek}, {Kuczewski}, {Kulkarni}, {Ku}, {Kurita}, {Lage}, {Lambert}, {Lange}, {Langton}, {Le Guillou}, {Levine}, {Liang}, {Lim}, {Lintott}, {Long}, {Lopez}, {Lotz}, {Lupton}, {Lust}, {MacArthur}, {Mahabal},
  {Mandelbaum}, {Markiewicz}, {Marsh}, {Marshall}, {Marshall}, {May}, {McKercher}, {McQueen}, {Meyers}, {Migliore}, {Miller}, {Mills}, {Miraval}, {Moeyens}, {Moolekamp}, {Monet}, {Moniez}, {Monkewitz}, {Montgomery}, {Morrison}, {Mueller}, {Muller}, {Mu{\~n}oz Arancibia}, {Neill}, {Newbry}, {Nief}, {Nomerotski}, {Nordby}, {O'Connor}, {Oliver}, {Olivier}, {Olsen}, {O'Mullane}, {Ortiz}, {Osier}, {Owen}, {Pain}, {Palecek}, {Parejko}, {Parsons}, {Pease}, {Peterson}, {Peterson}, {Petravick}, {Libby Petrick}, {Petry}, {Pierfederici}, {Pietrowicz}, {Pike}, {Pinto}, {Plante}, {Plate}, {Plutchak}, {Price}, {Prouza}, {Radeka}, {Rajagopal}, {Rasmussen}, {Regnault}, {Reil}, {Reiss}, {Reuter}, {Ridgway}, {Riot}, {Ritz}, {Robinson}, {Roby}, {Roodman}, {Rosing}, {Roucelle}, {Rumore}, {Russo}, {Saha}, {Sassolas}, {Schalk}, {Schellart}, {Schindler}, {Schmidt}, {Schneider}, {Schneider}, {Schoening}, {Schumacher}, {Schwamb}, {Sebag}, {Selvy}, {Sembroski}, {Seppala}, {Serio}, {Serrano}, {Shaw}, {Shipsey}, {Sick}, {Silvestri},
  {Slater}, {Smith}, {Smith}, {Sobhani}, {Soldahl}, {Storrie-Lombardi}, {Stover}, {Strauss}, {Street}, {Stubbs}, {Sullivan}, {Sweeney}, {Swinbank}, {Szalay}, {Takacs}, {Tether}, {Thaler}, {Thayer}, {Thomas}, {Thornton}, {Thukral}, {Tice}, {Trilling}, {Turri}, {Van Berg}, {Vanden Berk}, {Vetter}, {Virieux}, {Vucina}, {Wahl}, {Walkowicz}, {Walsh}, {Walter}, {Wang}, {Wang}, {Warner}, {Wiecha}, {Willman}, {Winters}, {Wittman}, {Wolff}, {Wood-Vasey}, {Wu}, {Xin}, {Yoachim}, \& {Zhan}}]{2019ApJ...873..111I}
{Ivezi{\'c}}, {\v{Z}}., {Kahn}, S.~M., {Tyson}, J.~A., {et~al.} 2019, \apj, 873, 111, \dodoi{10.3847/1538-4357/ab042c}

\bibitem[{{Jolicoeur} {et~al.}(2023){Jolicoeur}, {Maartens}, \& {Dlamini}}]{2023EPJC...83..320J}
{Jolicoeur}, S., {Maartens}, R., \& {Dlamini}, S. 2023, European Physical Journal C, 83, 320, \dodoi{10.1140/epjc/s10052-023-11482-2}

\bibitem[{{Komatsu} \& {Spergel}(2001)}]{2001PhRvD..63f3002K}
{Komatsu}, E., \& {Spergel}, D.~N. 2001, \prd, 63, 063002, \dodoi{10.1103/PhysRevD.63.063002}

\bibitem[{{Komatsu} {et~al.}(2003){Komatsu}, {Kogut}, {Nolta}, {Bennett}, {Halpern}, {Hinshaw}, {Jarosik}, {Limon}, {Meyer}, {Page}, {Spergel}, {Tucker}, {Verde}, {Wollack}, \& {Wright}}]{2003ApJS..148..119K}
{Komatsu}, E., {Kogut}, A., {Nolta}, M.~R., {et~al.} 2003, \apjs, 148, 119, \dodoi{10.1086/377220}

\bibitem[{{Leistedt} \& {Peiris}(2014)}]{2014MNRAS.444....2L}
{Leistedt}, B., \& {Peiris}, H.~V. 2014, \mnras, 444, 2, \dodoi{10.1093/mnras/stu1439}

\bibitem[{{Lewis} {et~al.}(2000){Lewis}, {Challinor}, \& {Lasenby}}]{2000ApJ...538..473L}
{Lewis}, A., {Challinor}, A., \& {Lasenby}, A. 2000, \apj, 538, 473, \dodoi{10.1086/309179}

\bibitem[{{Linde}(1982)}]{1982PhLB..108..389L}
{Linde}, A.~D. 1982, Physics Letters B, 108, 389, \dodoi{10.1016/0370-2693(82)91219-9}

\bibitem[{{Linde}(1983)}]{1983PhLB..129..177L}
---. 1983, Physics Letters B, 129, 177, \dodoi{10.1016/0370-2693(83)90837-7}

\bibitem[{{LSST Science Collaboration} {et~al.}(2009){LSST Science Collaboration}, {Abell}, {Allison}, {Anderson}, {Andrew}, {Angel}, {Armus}, {Arnett}, {Asztalos}, {Axelrod}, {Bailey}, {Ballantyne}, {Bankert}, {Barkhouse}, {Barr}, {Barrientos}, {Barth}, {Bartlett}, {Becker}, {Becla}, {Beers}, {Bernstein}, {Biswas}, {Blanton}, {Bloom}, {Bochanski}, {Boeshaar}, {Borne}, {Bradac}, {Brandt}, {Bridge}, {Brown}, {Brunner}, {Bullock}, {Burgasser}, {Burge}, {Burke}, {Cargile}, {Chandrasekharan}, {Chartas}, {Chesley}, {Chu}, {Cinabro}, {Claire}, {Claver}, {Clowe}, {Connolly}, {Cook}, {Cooke}, {Cooray}, {Covey}, {Culliton}, {de Jong}, {de Vries}, {Debattista}, {Delgado}, {Dell'Antonio}, {Dhital}, {Di Stefano}, {Dickinson}, {Dilday}, {Djorgovski}, {Dobler}, {Donalek}, {Dubois-Felsmann}, {Durech}, {Eliasdottir}, {Eracleous}, {Eyer}, {Falco}, {Fan}, {Fassnacht}, {Ferguson}, {Fernandez}, {Fields}, {Finkbeiner}, {Figueroa}, {Fox}, {Francke}, {Frank}, {Frieman}, {Fromenteau}, {Furqan}, {Galaz}, {Gal-Yam}, {Garnavich},
  {Gawiser}, {Geary}, {Gee}, {Gibson}, {Gilmore}, {Grace}, {Green}, {Gressler}, {Grillmair}, {Habib}, {Haggerty}, {Hamuy}, {Harris}, {Hawley}, {Heavens}, {Hebb}, {Henry}, {Hileman}, {Hilton}, {Hoadley}, {Holberg}, {Holman}, {Howell}, {Infante}, {Ivezic}, {Jacoby}, {Jain}, {R}, {Jedicke}, {Jee}, {Garrett Jernigan}, {Jha}, {Johnston}, {Jones}, {Juric}, {Kaasalainen}, {Styliani}, {Kafka}, {Kahn}, {Kaib}, {Kalirai}, {Kantor}, {Kasliwal}, {Keeton}, {Kessler}, {Knezevic}, {Kowalski}, {Krabbendam}, {Krughoff}, {Kulkarni}, {Kuhlman}, {Lacy}, {Lepine}, {Liang}, {Lien}, {Lira}, {Long}, {Lorenz}, {Lotz}, {Lupton}, {Lutz}, {Macri}, {Mahabal}, {Mandelbaum}, {Marshall}, {May}, {McGehee}, {Meadows}, {Meert}, {Milani}, {Miller}, {Miller}, {Mills}, {Minniti}, {Monet}, {Mukadam}, {Nakar}, {Neill}, {Newman}, {Nikolaev}, {Nordby}, {O'Connor}, {Oguri}, {Oliver}, {Olivier}, {Olsen}, {Olsen}, {Olszewski}, {Oluseyi}, {Padilla}, {Parker}, {Pepper}, {Peterson}, {Petry}, {Pinto}, {Pizagno}, {Popescu}, {Prsa}, {Radcka}, {Raddick},
  {Rasmussen}, {Rau}, {Rho}, {Rhoads}, {Richards}, {Ridgway}, {Robertson}, {Roskar}, {Saha}, {Sarajedini}, {Scannapieco}, {Schalk}, {Schindler}, {Schmidt}, {Schmidt}, {Schneider}, {Schumacher}, {Scranton}, {Sebag}, {Seppala}, {Shemmer}, {Simon}, {Sivertz}, {Smith}, {Allyn Smith}, {Smith}, {Spitz}, {Stanford}, {Stassun}, {Strader}, {Strauss}, {Stubbs}, {Sweeney}, {Szalay}, {Szkody}, {Takada}, {Thorman}, {Trilling}, {Trimble}, {Tyson}, {Van Berg}, {Vanden Berk}, {VanderPlas}, {Verde}, {Vrsnak}, {Walkowicz}, {Wandelt}, {Wang}, {Wang}, {Warner}, {Wechsler}, {West}, {Wiecha}, {Williams}, {Willman}, {Wittman}, {Wolff}, {Wood-Vasey}, {Wozniak}, {Young}, {Zentner}, \& {Zhan}}]{2009arXiv0912.0201L}
{LSST Science Collaboration}, {Abell}, P.~A., {Allison}, J., {et~al.} 2009, arXiv e-prints, arXiv:0912.0201, \dodoi{10.48550/arXiv.0912.0201}

\bibitem[{{Lyth} {et~al.}(2003){Lyth}, {Ungarelli}, \& {Wands}}]{2003PhRvD..67b3503L}
{Lyth}, D.~H., {Ungarelli}, C., \& {Wands}, D. 2003, \prd, 67, 023503, \dodoi{10.1103/PhysRevD.67.023503}

\bibitem[{{Maldacena}(2003)}]{2003JHEP...05..013M}
{Maldacena}, J. 2003, Journal of High Energy Physics, 2003, 013, \dodoi{10.1088/1126-6708/2003/05/013}

\bibitem[{{Moradinezhad Dizgah} {et~al.}(2021){Moradinezhad Dizgah}, {Biagetti}, {Sefusatti}, {Desjacques}, \& {Nore{\~n}a}}]{2021JCAP...05..015M}
{Moradinezhad Dizgah}, A., {Biagetti}, M., {Sefusatti}, E., {Desjacques}, V., \& {Nore{\~n}a}, J. 2021, \jcap, 2021, 015, \dodoi{10.1088/1475-7516/2021/05/015}

\bibitem[{{Mueller} {et~al.}(2022){Mueller}, {Rezaie}, {Percival}, {Ross}, {Ruggeri}, {Seo}, {Gil-Mar{\'\i}n}, {Bautista}, {Brownstein}, {Dawson}, {de la Macorra}, {Palanque-Delabrouille}, {Rossi}, {Schneider}, \& {Y{\'e}che}}]{2022MNRAS.514.3396M}
{Mueller}, E.-M., {Rezaie}, M., {Percival}, W.~J., {et~al.} 2022, \mnras, 514, 3396, \dodoi{10.1093/mnras/stac812}

\bibitem[{{Planck Collaboration} {et~al.}(2020){Planck Collaboration}, {Akrami}, {Arroja}, {Ashdown}, {Aumont}, {Baccigalupi}, {Ballardini}, {Banday}, {Barreiro}, {Bartolo}, {Basak}, {Benabed}, {Bernard}, {Bersanelli}, {Bielewicz}, {Bond}, {Borrill}, {Bouchet}, {Bucher}, {Burigana}, {Butler}, {Calabrese}, {Cardoso}, {Casaponsa}, {Challinor}, {Chiang}, {Colombo}, {Combet}, {Crill}, {Cuttaia}, {de Bernardis}, {de Rosa}, {de Zotti}, {Delabrouille}, {Delouis}, {Di Valentino}, {Diego}, {Dor{\'e}}, {Douspis}, {Ducout}, {Dupac}, {Dusini}, {Efstathiou}, {Elsner}, {En{\ss}lin}, {Eriksen}, {Fantaye}, {Fergusson}, {Fernandez-Cobos}, {Finelli}, {Frailis}, {Fraisse}, {Franceschi}, {Frolov}, {Galeotta}, {Galli}, {Ganga}, {G{\'e}nova-Santos}, {Gerbino}, {Gonz{\'a}lez-Nuevo}, {G{\'o}rski}, {Gratton}, {Gruppuso}, {Gudmundsson}, {Hamann}, {Handley}, {Hansen}, {Herranz}, {Hivon}, {Huang}, {Jaffe}, {Jones}, {Jung}, {Keih{\"a}nen}, {Keskitalo}, {Kiiveri}, {Kim}, {Krachmalnicoff}, {Kunz}, {Kurki-Suonio}, {Lamarre}, {Lasenby},
  {Lattanzi}, {Lawrence}, {Le Jeune}, {Levrier}, {Lewis}, {Liguori}, {Lilje}, {Lindholm}, {L{\'o}pez-Caniego}, {Ma}, {Mac{\'\i}as-P{\'e}rez}, {Maggio}, {Maino}, {Mandolesi}, {Marcos-Caballero}, {Maris}, {Martin}, {Mart{\'\i}nez-Gonz{\'a}lez}, {Matarrese}, {Mauri}, {McEwen}, {Meerburg}, {Meinhold}, {Melchiorri}, {Mennella}, {Migliaccio}, {Miville-Desch{\^e}nes}, {Molinari}, {Moneti}, {Montier}, {Morgante}, {Moss}, {M{\"u}nchmeyer}, {Natoli}, {Oppizzi}, {Pagano}, {Paoletti}, {Partridge}, {Patanchon}, {Perrotta}, {Pettorino}, {Piacentini}, {Polenta}, {Puget}, {Rachen}, {Racine}, {Reinecke}, {Remazeilles}, {Renzi}, {Rocha}, {Rubi{\~n}o-Mart{\'\i}n}, {Ruiz-Granados}, {Salvati}, {Savelainen}, {Scott}, {Shellard}, {Shiraishi}, {Sirignano}, {Sirri}, {Smith}, {Spencer}, {Stanco}, {Sunyaev}, {Suur-Uski}, {Tauber}, {Tavagnacco}, {Tenti}, {Toffolatti}, {Tomasi}, {Trombetti}, {Valiviita}, {Van Tent}, {Vielva}, {Villa}, {Vittorio}, {Wandelt}, {Wehus}, {Zacchei}, \& {Zonca}}]{2020A&A...641A...9P}
{Planck Collaboration}, {Akrami}, Y., {Arroja}, F., {et~al.} 2020, \aap, 641, A9, \dodoi{10.1051/0004-6361/201935891}

\bibitem[{{Pullen} \& {Hirata}(2013)}]{2013PASP..125..705P}
{Pullen}, A.~R., \& {Hirata}, C.~M. 2013, \pasp, 125, 705, \dodoi{10.1086/671189}

\bibitem[{{Rezaie} {et~al.}(2021){Rezaie}, {Ross}, {Seo}, {Mueller}, {Percival}, {Merz}, {Katebi}, {Bunescu}, {Bautista}, {Brownstein}, {Burtin}, {Dawson}, {Gil-Mar{\'\i}n}, {Hou}, {Lyke}, {de la Macorra}, {Rossi}, {Schneider}, {Zarrouk}, \& {Zhao}}]{2021MNRAS.506.3439R}
{Rezaie}, M., {Ross}, A.~J., {Seo}, H.-J., {et~al.} 2021, \mnras, 506, 3439, \dodoi{10.1093/mnras/stab1730}

\bibitem[{{Rezaie} {et~al.}(2024){Rezaie}, {Ross}, {Seo}, {Kong}, {Porredon}, {Samushia}, {Chaussidon}, {Krolewski}, {de Mattia}, {Beutler}, {Aguilar}, {Ahlen}, {Alam}, {Avila}, {Bahr-Kalus}, {Bermejo-Climent}, {Brooks}, {Claybaugh}, {Cole}, {Dawson}, {de la Macorra}, {Doel}, {Font-Ribera}, {Forero-Romero}, {Gontcho}, {Guy}, {Honscheid}, {Huterer}, {Kisner}, {Landriau}, {Levi}, {Manera}, {Meisner}, {Miquel}, {Mueller}, {Myers}, {Newman}, {Nie}, {Palanque-Delabrouille}, {Percival}, {Poppett}, {Rossi}, {Sanchez}, {Schubnell}, {Tarl{\'e}}, {Weaver}, {Y{\`e}che}, {Zhou}, \& {Zou}}]{2024MNRAS.532.1902R}
---. 2024, \mnras, 532, 1902, \dodoi{10.1093/mnras/stae886}

\bibitem[{{Ross} {et~al.}(2013){Ross}, {Percival}, {Carnero}, {Zhao}, {Manera}, {Raccanelli}, {Aubourg}, {Bizyaev}, {Brewington}, {Brinkmann}, {Brownstein}, {Cuesta}, {da Costa}, {Eisenstein}, {Ebelke}, {Guo}, {Hamilton}, {Maga{\~n}a}, {Malanushenko}, {Malanushenko}, {Maraston}, {Montesano}, {Nichol}, {Oravetz}, {Pan}, {Prada}, {S{\'a}nchez}, {Samushia}, {Schlegel}, {Schneider}, {Seo}, {Sheldon}, {Simmons}, {Snedden}, {Swanson}, {Thomas}, {Tinker}, {Tojeiro}, \& {Zehavi}}]{2013MNRAS.428.1116R}
{Ross}, A.~J., {Percival}, W.~J., {Carnero}, A., {et~al.} 2013, \mnras, 428, 1116, \dodoi{10.1093/mnras/sts094}

\bibitem[{{Saraf} {et~al.}(2024){Saraf}, {Bielewicz}, \& {Chodorowski}}]{2024A&A...690A.338S}
{Saraf}, C.~S., {Bielewicz}, P., \& {Chodorowski}, M. 2024, \aap, 690, A338, \dodoi{10.1051/0004-6361/202450749}

\bibitem[{{Sato}(1981)}]{1981MNRAS.195..467S}
{Sato}, K. 1981, \mnras, 195, 467, \dodoi{10.1093/mnras/195.3.467}

\bibitem[{{Schmittfull} \& {Seljak}(2018)}]{2018PhRvD..97l3540S}
{Schmittfull}, M., \& {Seljak}, U. 2018, \prd, 97, 123540, \dodoi{10.1103/PhysRevD.97.123540}

\bibitem[{{Shekhar Saraf} \& {Bielewicz}(2024)}]{2024A&A...687A.150S}
{Shekhar Saraf}, C., \& {Bielewicz}, P. 2024, \aap, 687, A150, \dodoi{10.1051/0004-6361/202348732}

\bibitem[{{Slosar} {et~al.}(2008){Slosar}, {Hirata}, {Seljak}, {Ho}, \& {Padmanabhan}}]{2008JCAP...08..031S}
{Slosar}, A., {Hirata}, C., {Seljak}, U., {Ho}, S., \& {Padmanabhan}, N. 2008, \jcap, 2008, 031, \dodoi{10.1088/1475-7516/2008/08/031}

\bibitem[{{Starobinsky}(1980)}]{1980PhLB...91...99S}
{Starobinsky}, A.~A. 1980, Physics Letters B, 91, 99, \dodoi{10.1016/0370-2693(80)90670-X}

\bibitem[{{Sullivan} {et~al.}(2023){Sullivan}, {Prijon}, \& {Seljak}}]{2023JCAP...08..004S}
{Sullivan}, J.~M., {Prijon}, T., \& {Seljak}, U. 2023, \jcap, 2023, 004, \dodoi{10.1088/1475-7516/2023/08/004}

\bibitem[{{Tegmark} {et~al.}(2004){Tegmark}, {Strauss}, {Blanton}, {Abazajian}, {Dodelson}, {Sandvik}, {Wang}, {Weinberg}, {Zehavi}, {Bahcall}, {Hoyle}, {Schlegel}, {Scoccimarro}, {Vogeley}, {Berlind}, {Budavari}, {Connolly}, {Eisenstein}, {Finkbeiner}, {Frieman}, {Gunn}, {Hui}, {Jain}, {Johnston}, {Kent}, {Lin}, {Nakajima}, {Nichol}, {Ostriker}, {Pope}, {Scranton}, {Seljak}, {Sheth}, {Stebbins}, {Szalay}, {Szapudi}, {Xu}, {Annis}, {Brinkmann}, {Burles}, {Castander}, {Csabai}, {Loveday}, {Doi}, {Fukugita}, {Gillespie}, {Hennessy}, {Hogg}, {Ivezi{\'c}}, {Knapp}, {Lamb}, {Lee}, {Lupton}, {McKay}, {Kunszt}, {Munn}, {O'Connell}, {Peoples}, {Pier}, {Richmond}, {Rockosi}, {Schneider}, {Stoughton}, {Tucker}, {vanden Berk}, {Yanny}, \& {York}}]{2004PhRvD..69j3501T}
{Tegmark}, M., {Strauss}, M.~A., {Blanton}, M.~R., {et~al.} 2004, \prd, 69, 103501, \dodoi{10.1103/PhysRevD.69.103501}

\bibitem[{{Tessore} {et~al.}(2023){Tessore}, {Loureiro}, {Joachimi}, {von Wietersheim-Kramsta}, \& {Jeffrey}}]{2023OJAp....6E..11T}
{Tessore}, N., {Loureiro}, A., {Joachimi}, B., {von Wietersheim-Kramsta}, M., \& {Jeffrey}, N. 2023, The Open Journal of Astrophysics, 6, 11, \dodoi{10.21105/astro.2302.01942}

\bibitem[{{Zaldarriaga}(2004)}]{2004PhRvD..69d3508Z}
{Zaldarriaga}, M. 2004, \prd, 69, 043508, \dodoi{10.1103/PhysRevD.69.043508}

\bibitem[{{Zhang} {et~al.}(2010){Zhang}, {Pen}, \& {Bernstein}}]{2010MNRAS.405..359Z}
{Zhang}, P., {Pen}, U.-L., \& {Bernstein}, G. 2010, \mnras, 405, 359, \dodoi{10.1111/j.1365-2966.2010.16445.x}

\end{thebibliography}
\bibliographystyle{aasjournal}



\end{document}